\documentclass[trackchanges,twocolumn,twocolappendix]{aastex7}
\usepackage{amsmath}
\usepackage{float}

\begin{document}

\title{Particle Acceleration in Magnetized Shear-Driven Turbulence}

\author[0009-0009-1522-9581]{Mingxuan Liu}
\affiliation{Department of Astronomy, University of Michigan, Ann Arbor, MI 48109, USA}
\email[show]{mgxnliu@umich.edu}

\author[0009-0002-2669-9908]{Mateusz Ruszkowski}
\affiliation{Department of Astronomy, University of Michigan, Ann Arbor, MI 48109, USA}
\email{mateuszr@umich.edu}

\author[0000-0003-4821-713X]{Ellen Zweibel}
\affiliation{Department of Astronomy, University of Wisconsin–Madison, Madison, WI 53706, USA}
\email{zweibel@astro.wisc.edu}

\author{Xiaochen Sun}
\affiliation{Department of Astrophysical Sciences, Princeton University, Princeton, NJ 08544, USA}
\email{sun.xiaochen@princeton.edu}

\author[0000-0003-0939-8775]{Damiano Caprioli}
\affiliation{Department of Astronomy and Astrophysics, University of Chicago, Chicago, IL 60637, USA}
\email{caprioli@uchicago.edu}

\author[0009-0003-0137-3116]{Naixin Liang}
\affiliation{Department of Physics, University of California, Santa Barbara, CA 93106, USA}
\email{naixin@ucsb.edu}

\author[0000-0002-1013-4657]{S. Peng Oh}
\affiliation{Department of Physics, University of California, Santa Barbara, CA 93106, USA}
\email{peng@physics.ucsb.edu}

\author[0000-0001-9179-9054]{Anatoly Spitkovsky}
\affiliation{Department of Astrophysical Sciences, Princeton University, Princeton, NJ 08544, USA}
\email{anatoly@princeton.edu}

\begin{abstract}
Shear flows, ubiquitous in space and astrophysical plasmas, can accelerate particles through turbulence excited by the Kelvin-Helmholtz instability. We present the first numerical study of particle acceleration in sustained, subsonic, non-relativistic, and magnetized turbulence driven purely by velocity shear, including full particle backreaction. Using two-dimensional MHD-PIC simulations with an initially uniform flow-aligned magnetic field and external stirring force, we demonstrate that, in this subsonic regime, sustained particle acceleration requires continuously driven turbulence, whereas freely decaying turbulence rapidly depletes its energy reservoirs and halts the acceleration. The acceleration mechanism operates through the systematic distortion of gyro-orbits by turbulent electric fields: acceleration phases extend the particle trajectory along the electric force, increasing the energy gain, while deceleration phases shorten the trajectory, reducing the energy loss. This asymmetry produces net energy gain despite stochastic fluctuations, with the mean energy change scaling quadratically with shear velocity, characteristic of second-order Fermi acceleration. Initially monoenergetic particles develop substantial non-thermal tails after the turbulence onset. The energization of particles that repeatedly cross shear layers follows geometric Brownian motion, yielding a log-normal distribution. High-energy particles exhibit pitch-angle anisotropy, becoming preferentially perpendicular to the flow-aligned magnetic field as their gyroradii exceed the turbulent layer width. These results establish shear-driven turbulence as a viable particle acceleration mechanism, providing a general model for particle energization in shear flows.
\end{abstract}


\section{Introduction}
Shear flows are ubiquitous in space and astrophysical plasmas across a wide range of dynamical scales, from the solar wind-magnetosphere interface \citep{miura_simulation_1987, henri_nonlinear_2013, archer_magnetopause_2021, goodwill_nonlinear_2025}, to accretion flows around compact objects \citep{Subramanian_1999}, to relativistic outflows in gamma-ray bursts and active galactic nuclei \citep{rieger_introduction_2019}, to sloshing cold fronts in the intracluster medium \citep{roediger_kelvinhelmholtz_2013,walker_is_2017}.

When two fluids move relative to each other, the Kelvin-Helmholtz instability (KHI) can develop at their interface. The instability arises from Bernoulli's principle: small perturbations of the shear-layer interface cause local flow acceleration over the crests, reducing pressure and further amplifying the initial disturbance. This positive feedback drives the formation of periodic vortex roll-ups that eventually transition to turbulence. The resulting turbulence drives fluid mixing and momentum transfer across shear layers, which create conditions conducive to \textit{in situ} particle acceleration \citep{ferrari_magnetohydrodynamic_1979, rieger_turbulence_2021, wang_particle_2023}.

Particle acceleration in shear flows operates primarily through a second-order Fermi-like mechanism \citep{earl_cosmic-ray_1988, rieger_shear_2004, rieger_fermi_2007, webb_particle_2018, webb_particle_2019, rieger_introduction_2019, lemoine_generalized_2019}, an idea first explored by \citet{berezhko_kinetics_1981}. Particles gain energy by repeatedly crossing the shear layer and scattering off magnetic inhomogeneities embedded in the turbulence. Each crossing allows particles to sample the velocity difference across the shear, with turbulent scattering randomizing their pitch angles between encounters. This stochastic process leads to net energy gain despite individual scattering events being either favorable or unfavorable. The accelerated particles, in turn, affect the background flow dynamics. As they cross the shear layer, particles transfer momentum between fluids moving at different velocities, acting as an effective viscosity that damps the velocity shear \citep{earl_cosmic-ray_1988}. Therefore, particle acceleration in shear flows is crucial not only for understanding cosmic ray energization but also for quantifying how energetic particles modify the evolution of astrophysical shear flows.

Recent first-principles particle-in-cell (PIC) studies have established that magnetized turbulence can generate efficient non-thermal particle acceleration. \citet{comisso_particle_2018, comisso_interplay_2019} showed that particle acceleration in magnetically dominated relativistic turbulence proceeds through a two-stage process: initial injection by plasmoid-mediated reconnection at strong current sheets (dominated by parallel electric fields $E_\parallel$), followed by stochastic acceleration of high-energy particles via turbulent fluctuations (dominated by perpendicular fields $E_\perp$). \citet{wong_first-principles_2020} provided the first direct validation of the Fokker-Planck framework in driven magnetized turbulence with relativistic pair plasma, demonstrating that the energy diffusion coefficient scales as $D(\gamma) \propto \gamma^2$, which is consistent with second-order Fermi acceleration. Beyond the scattering acceleration, \citet{das_studying_2025} recently identified mirror acceleration as an alternative mechanism in strongly magnetized relativistic turbulence, finding that particles gain energy preferentially perpendicular to local magnetic fields due to the induced electric field $E_\perp$, with pitch angles becoming increasingly anisotropic at higher energies.

Despite these advances, direct numerical studies of particle acceleration in turbulence specifically excited by KHI remain limited. Recent PIC simulations have begun to address this gap, particularly in the context of relativistic jets. \citet{sironi_reconnection-driven_2021} demonstrated that combining velocity shear with out-of-plane magnetic shear in relativistic pair plasmas develops nonlinear KHI and generates kinetic-scale reconnection layers. They found that while magnetic reconnection provides initial particle injection at current sheets, the acceleration mechanism at higher energy is dominated by turbulent scattering, following a similar two-stage process reported by \citet{comisso_particle_2018, comisso_interplay_2019}. Building on this work, \citet{tsung_dissipation_2025} focused on the joint effects of velocity and magnetic shear by suppressing tearing modes. This approach isolates the Kelvin-Helmholtz and Drift-Kink instabilities (DKI) while minimizing the role of magnetic reconnection. They showed that the combination of KHI and DKI significantly modifies flow structures and enhances both energy dissipation and particle acceleration compared to KHI alone. However, both works studied decaying turbulence with relativistic pair plasma and included other mechanisms that modify the turbulence alongside KHI.

In this study, we investigate particle acceleration in a fundamentally different regime: subsonic, non-relativistic, and magnetized turbulence driven purely by velocity shear. To focus on the effects of shear-driven turbulence, we adopt the ideal MHD limit with a uniform magnetic field aligned with the flow direction, thereby precluding initial magnetic shear. Unlike previous studies such as \citet{comisso_particle_2018,comisso_interplay_2019} and \citet{sironi_reconnection-driven_2021}, where magnetic reconnection serves as an injection mechanism for particles to enter stochastic acceleration, we initialize the particles energetic enough to readily cross shear layers from the outset, thereby bypassing the injection problem. While current sheets can naturally form in 2D turbulence, their subsequent dissipation occurs at the grid scale due to numerical resistivity, which only affects the MHD fluid fields and does not directly energize the particles in our MHD-PIC framework. This configuration therefore isolates the role of KHI-driven turbulence in particle acceleration, which proceeds solely through the stochastic interactions with the turbulent motional electric field. Our simulations with such a setup are not only conceptual test cases, but also relevant to environments beyond relativistic jets, where cosmic ray acceleration has been studied, such as galaxy cluster turbulence \citep{brunetti_particle_2001,brunetti_compressible_2007}, accretion flows \citep{Subramanian_1999,kimura_stochastic_2016,sun_particle_2021}, protostellar jets \citep{padovani_protostars_2016,mendez-gallego_exploring_2025}, and planetary magnetosphere boundaries \citep{burgess_ion_2012,delamere_kelvinhelmholtz-related_2021}; see \citet{ruszkowski2023cosmic} for a recent review of the role of cosmic rays. We employ two-dimensional MHD-PIC simulations with a stirring force that maintains the shear profile against dissipation, enabling us to study particle acceleration in sustained, quasi-stationary turbulence.

This paper is organized as follows. Section \ref{methods} describes our numerical approach, including the MHD-PIC governing equations and simulation setup. Section \ref{results} presents our findings on turbulence characteristics, particle energy distributions, and the acceleration mechanism. Section \ref{discussion} discusses potential applications of our results to astrophysical environments and places these results in the context of related works. We summarize the findings and conclude the paper in Section \ref{conclusions}.

\section{Methods}\label{methods}
To study particle acceleration in fully developed turbulence, we performed 2D simulations using the magnetohydrodynamic-particle-in-cell (MHD-PIC) code \citep{sun_magnetohydrodynamic-particle--cell_2023} integrated into the ATHENA++ framework \citep{stone_athena_2020}. The MHD-PIC code treats energetic particles as in the standard PIC approach, while modeling the thermal particles as the background fluid described by the MHD equations. This approach substantially alleviates the problem of extreme scale separation among plasma skin depth, thermal particle gyroradius, and the global system size, thus greatly reducing the computational cost of simulations while retaining all kinetic physics.

\subsection{Governing Equations}\label{equations}
In our simulation, both the kinetic particles and background fluids are spatially confined in the $x$--$y$ plane, but we keep all three components ($\boldsymbol{\hat{x}}, \boldsymbol{\hat{y}},\boldsymbol{\hat{z}}$) of their vector quantities. Each simulation particle carries both the 2D position information $\boldsymbol{x}$ and mass-normalized momentum information $(\boldsymbol{p}/m)$. The particle kinetic equations are:
\begin{align}
\frac{\mathrm{d} \boldsymbol{x}}{\mathrm{d} t} &= \boldsymbol{v}, \label{eq:eom_start} \\
\frac{\mathrm{d}(\boldsymbol{p} / m)}{\mathrm{d} t} &= \left(\frac{q}{mc}\right)(c \boldsymbol{E}+\boldsymbol{v} \times \boldsymbol{B}), \\
\boldsymbol{v} &\equiv \frac{(\boldsymbol{p} / m)}{\gamma}, \quad \gamma \equiv \sqrt{1+(\boldsymbol{p} / m)^{2} / \mathbb{C}^{2}}, \label{eq:eom_end}
\end{align}
where $q/(mc)$, $\boldsymbol{v}$, $\gamma$, and $\mathbb{C}$ represent the particle charge-to-mass ratio, particle velocity, Lorentz factor, and the numerical speed of light, respectively. In non-relativistic MHD, no physical scale is set by the speed of light, and the factors of $c$ above act only as unit conversions through the combinations $c\boldsymbol{E}$ and $q/(mc)$, into which they can be absorbed \citep{bai_magnetohydrodynamic-particle--cell_2015, sun_magnetohydrodynamic-particle--cell_2023}. The particles, in contrast, follow relativistic dynamics, so the speed of light entering the Lorentz factor must be specified explicitly; because the fluid equations provide no light-speed scale, $\mathbb{C}$ is a numerical parameter of the MHD-PIC method, not a physical constant \citep{bai_magnetohydrodynamic-particle--cell_2015}. For consistency with the MHD background, its value must greatly exceed all characteristic fluid speeds, while excessively large values raise the computational cost \citep{bai_magnetohydrodynamic-particle--cell_2015, sun_magnetohydrodynamic-particle--cell_2023}. The particles exert a Lorentz force on the background fluids as backreaction. The MHD equations with particle backreaction are given as:
\begin{align}
\partial_{t} \rho &= -\nabla \cdot\left(\rho \boldsymbol{u}\right), \\
\partial_{t} \boldsymbol{B} &= -\nabla \times(c \boldsymbol{E}), \\
c \boldsymbol{E} &= -\boldsymbol{u} \times \boldsymbol{B},
\end{align}
\vspace{-\baselineskip}
\begin{equation}
\begin{aligned}
\partial_{t}\left(\rho \boldsymbol{u}\right)+\nabla \cdot\left(\rho \boldsymbol{u}^{T} \boldsymbol{u}-\boldsymbol{B}^{T} \boldsymbol{B}+\mathbb{P}\right) \\
=F_{\text{stir}} - \left(\frac{Q_{\mathrm{p}}}{c} c \boldsymbol{E}+\frac{\boldsymbol{j}_{\mathrm{p}}}{c} \times \boldsymbol{B}\right),
\end{aligned}
\end{equation}
where $\rho$ represents thermal ion density, $\boldsymbol{u}$ represents thermal ion velocity, $\mathbb{P} \equiv (\rho c_s^2 + B^2/2)\mathbb{I}$ with $\mathbb{I}$ being the identity tensor, and $c_s$ is the isothermal sound speed. $F_{\text{stir}}$ is the external driving force defined in Equation \ref{force}. $Q_{\mathrm{p}}$ and $\boldsymbol{j}_{\mathrm{p}}$ denote the particle charge density and current density, both as functions of the particle phase space distribution function $f(t, \boldsymbol{x}, \boldsymbol{p}/m)$:
\begin{equation}
    \begin{aligned}
        \frac{Q_p}{c} &\equiv \left(\frac{q}{mc}\right) m \int{f d^3\boldsymbol{p}} = \left(\frac{q}{mc}\right)\rho_p, \\
        \frac{\boldsymbol{j}_p}{c} &\equiv \left(\frac{q}{mc}\right)m \int{\boldsymbol{v}f d^3\boldsymbol{p}},
    \end{aligned}
\end{equation}
where $m$ represents the individual particle mass, and $\rho_p$ represents the particle mass density.

To integrate these governing equations, we utilized the two-staged van Leer time integrator \citep{stone_simple_2009} to solve MHD equations and the Boris pusher \citep{boris_proceedings_1973} to integrate kinetic particles, achieving a second-order temporal accuracy \citep{sun_magnetohydrodynamic-particle--cell_2023}.

\subsection{Numerical Setup} \label{setup}
In this work, we consider a non-relativistic ($\mathbb{C} = 50\,U_0$), subsonic ($c_s = 5\,U_0$), and super-Alfv\'enic ($U_A = 0.1\,U_0$) flow, where $U_0$ is the characteristic fluid velocity. This choice places $\mathbb{C}$ well above every characteristic fluid speed ($\mathbb{C} = 10\,c_s = 500\,U_A$), as the MHD-PIC formulation requires. In addition, the background fluid has a uniform initial mass density $\rho_0$. We adopt the ideal MHD limit by neglecting dissipation terms such as viscosity (i.e., $\nu = 0$) and resistivity (i.e., $\eta = 0$). Although intermittent current sheets can form as magnetic field lines are stretched and folded by the flow, their localized numerical dissipation only contributes to the energy cascade and does not provide direct particle energization. As a result, particle energization occurs solely through interactions with the electromagnetic fields generated by the shear-driven turbulence. We further adopt an isothermal equation of state \citep{bai_magnetohydrodynamic-particle--cell_2015, sun_magnetohydrodynamic-particle--cell_2023}, under which the heat generated by turbulent dissipation is removed rather than accumulated. Together with the continuous driving described below, this closure holds the background thermodynamic state fixed and allows the turbulence to reach the statistically stationary state characterized in Section~\ref{turbulence}, enabling a clean measurement of long-term particle energization. Because the flow is subsonic and the plasma $\beta$ is high, the velocity and magnetic fields that generate the motional electric fields are largely insensitive to the thermal state. We return to this idealization in Section~\ref{discussion}.

\begin{figure}
    \centering
    \includegraphics[width=1.0\linewidth]{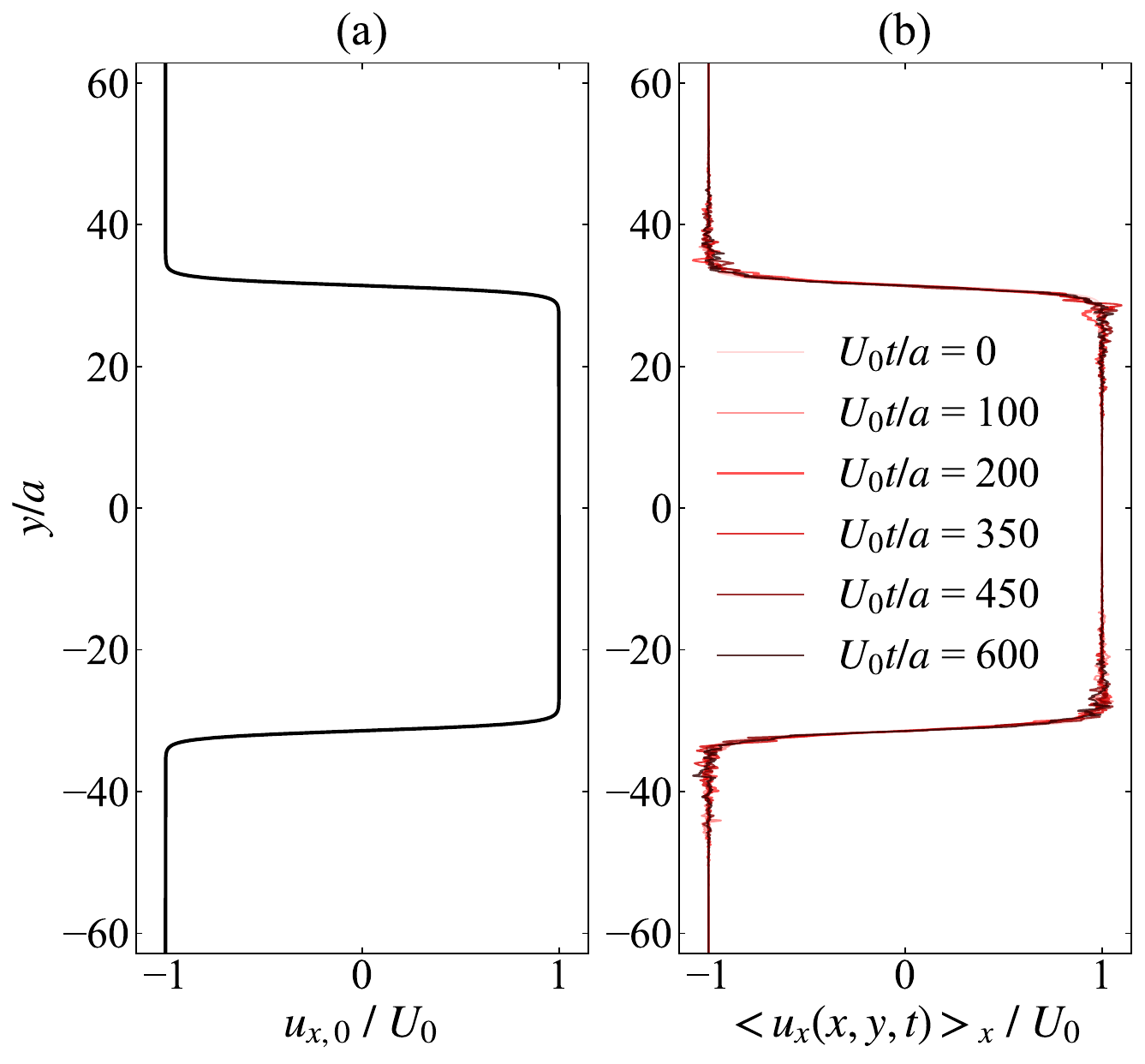}
    \caption{(a) Initial shear profile of background plasma's velocity field, as a function of y positions. (b) $X$-averaged mean flow profile at different times, ranging from $U_0 t/a = 0$ to $600$. Both profiles are normalized by the characteristic flow velocity $U_0$. Fluctuations near the velocity transition region are caused by turbulent eddies.}
    \label{fig:shear_profiles}
\end{figure}

To excite the Kelvin-Helmholtz instability, we initialized a double-layer setup, with each shear layer following a hyperbolic tangent velocity profile (see Figure~\ref{fig:shear_profiles}(a)):
\begin{equation}\label{shear_velocity}
    u_{x,0}(y)=U_{0}\left[1-\tanh \left(\frac{y-y_{1}}{a}\right)+\tanh \left(\frac{y-y_{2}}{a}\right)\right],
\end{equation}
where $a$ is the layer half-width that controls the steepness of the velocity transition; $y_1$ and $y_2$ are the positions for lower and upper shear layers, respectively. In this setup, the shear flow moves in the $x$-direction and the shear velocity varies along the $y$-direction. 

We also imposed an initially uniform and flow-aligned magnetic field $\boldsymbol{B}=B_0 \,\boldsymbol{\hat{x}}$, independent of the shear profile, so that particles predominantly gyrate in the $y$--$z$ plane and repeatedly cross shear layers. By construction, the motional electric field $\boldsymbol{E} = -\boldsymbol{u}\times\boldsymbol{B}/c$ is initially zero and does not energize the particles until the onset of the Kelvin-Helmholtz instability.

Furthermore, to maintain the shear flow against energy dissipation, we applied an external body force (namely, the stirring force) in the $x$-direction that adaptively drives the mean flow profile toward the initial shear profile \citep{tripathi_near-cancellation_2022}:
\begin{equation}\label{force}
    F_{\text{stir}}(y, t)=\rho(x, y, t) \left(\frac{u_{x,0}(y) - \left\langle u_{x}(x, y, t)\right\rangle_{x}}{\tau_r}\right),
\end{equation}
where $\rho(x, y, t)$ is the background fluid density, $\tau_r=0.5\,a/U_0$ is the profile relaxation timescale, and $\left\langle u_{x}(x, y, t)\right\rangle_{x}$ represents the $x$-averaged mean flow profile. Common in astrophysical shear flows \citep{witzke_evolution_2016}, the stirring force serves as a proxy for the continuous energy injection provided by large-scale drivers, such as the gravitational potential well in accretion flows \citep{1998RvMP...70....1B,2002apa..book.....F,Li_2026} or the central engine in AGN jets \citep{1977MNRAS.179..433B,Giroletti_2004,2016A&A...595A..54M}, which operate on timescales far exceeding the local eddy turnover time. By maintaining the shear profile (see Figure~\ref{fig:shear_profiles}(b)), our setup mimics the quasi-stationary turbulence in realistic astrophysical environments, where the flow is continuously replenished against dissipation.

The simulation domain has periodic boundary conditions in all directions. It spans $L_x = 10\pi\, a$ in the $x$-direction with 1024 cells, $L_y = 40\pi\, a$ in the $y$-direction with 4096 cells, and $L_z = 1\,a$ in the $z$-direction with 1 cell. Two shear layers are located at $y_{1,2} = \{-L_y/4,\ +L_y/4\}$, each resolved by $\sim 66$ grid cells. 

Lastly, $\sim 7\times10^7$ single-species particles are uniformly seeded throughout the domain, yielding 16 particles per cell with mass density $\rho_{p} = 5 \times 10^{-4}\, \rho_0$, where $\rho_0$ is the initial mass density of the background fluid. Throughout this work, the particle energy density is the volume-averaged particle kinetic energy, $\varepsilon_p \equiv \sum_j (\gamma_j - 1)\, m\, \mathbb{C}^2 / V$, where the sum runs over all particles in the simulation domain of volume $V$ and $\gamma_j$ is the Lorentz factor of particle $j$ (Equation~\ref{eq:eom_end}); the corresponding fluid kinetic and magnetic energy densities are $\varepsilon_k \equiv \left\langle \rho\, |\boldsymbol{u}|^2 \right\rangle / 2$ and $\varepsilon_m \equiv \left\langle |\boldsymbol{B}|^2 \right\rangle / 2$, with $\left\langle \cdot \right\rangle$ denoting the volume average. The particle mass density is chosen such that $\varepsilon_p$ is initially comparable to $\varepsilon_k$, enabling particle backreaction on the flow dynamics; this choice mirrors the approximate equipartition between the cosmic ray, thermal, magnetic, and turbulent energy densities observed in the interstellar medium \citep{boulares_galactic_1990, ferriere_interstellar_2001, ruszkowski2023cosmic}. The initial particle energy distribution is described by a Dirac delta function centered at $\gamma_0 = \sqrt{2}$ in the local fluid rest frame, corresponding to an initial mass-normalized momentum $(p/m)_0 = \mathbb{C}$. With these parameters, the initial energy densities are $\varepsilon_p \approx 0.52\,\rho_0 U_0^2$ and $\varepsilon_k \approx 0.48\,\rho_0 U_0^2$, so the particle energy density slightly exceeds the fluid kinetic energy density from the outset and grows further as particles are accelerated (Figure~\ref{fig:steady_state}). The initial gyroradius is defined as the ensemble average $r_{\rm c,0} \equiv \left\langle p_{\perp}/m \right\rangle / \left[ (q/mc)\, B_0 \right]$, where $p_{\perp}$ is the momentum component perpendicular to the magnetic field, so that $r_{\rm c,0}$ accounts for the particle pitch angles. The charge-to-mass ratio $q/(mc)$ is chosen such that the initial gyroradii exceed the shear layer width: $r_{\rm c,0} \gtrsim 2\,a \equiv \Delta$. This setup allows particles to readily cross shear layers from the outset, effectively bypassing the injection problem. Our approach thus models the reacceleration phase relevant to environments where a pre-existing energetic particle population interacts with shear-driven turbulence. The initial particle velocity directions are randomly assigned in the local fluid rest frame so that the velocity distribution is isotropic with respect to the shear flow (see Appendix~\ref{details} for details). While our fiducial setup focuses on $r_{\rm c,0} \gtrsim \Delta$, we have verified that sustained particle acceleration also occurs for initial gyroradii down to $r_{\rm c,0} \simeq 0.2\,\Delta$ (see Appendix~\ref{extrapolation}), indicating that the acceleration mechanism operates across a wider range of initial conditions. Nevertheless, a complete treatment of bottom-up injection from the thermal pool would require fully kinetic or hybrid simulations that capture the microphysics at kinetic scales, which remains beyond the scope of this study.

\section{Results}\label{results}
In this section, we discuss the process of particle acceleration occurring through interactions with shear-driven turbulence. Specifically, in Section \ref{turbulence}, we provide a brief overview of the steady-state characteristics of shear-driven turbulence generated by the Kelvin-Helmholtz instability, as well as its temporal evolution under a driving force; in Section \ref{spectrum}, we show that particles can be consistently accelerated, forming non-thermal tails after the turbulence onset; in Section \ref{acceleration}, we characterize the microscopic acceleration mechanism driven by gyro-orbit distortion and demonstrate that the resulting stochastic energization follows geometric Brownian motion.

\subsection{Shear-Driven Turbulence} \label{turbulence}
The initially laminar shear flow configuration described in Section \ref{setup} is susceptible to the Kelvin-Helmholtz instability due to the velocity gradient across the shear layer. However, a uniform magnetic field aligned with the flow direction can stabilize the KHI through magnetic tension, an effect first demonstrated in the linear analysis of \citet{chandrasekhar_hydrodynamic_1961}. The instability is suppressed when the Alfv\'en Mach number $M_A$ falls below a critical threshold of approximately 1--2 \citep{ryu_magnetohydrodynamic_2000, palotti_evolution_2008,  fraser_impact_2021}. Our simulations operate in the weak-field regime with $M_A = 10$, well above this critical value, ensuring that the instability develops with only a slight reduction in growth rate compared to the hydrodynamic case.

\begin{figure}
    \centering
    \includegraphics[width=1.0\linewidth]{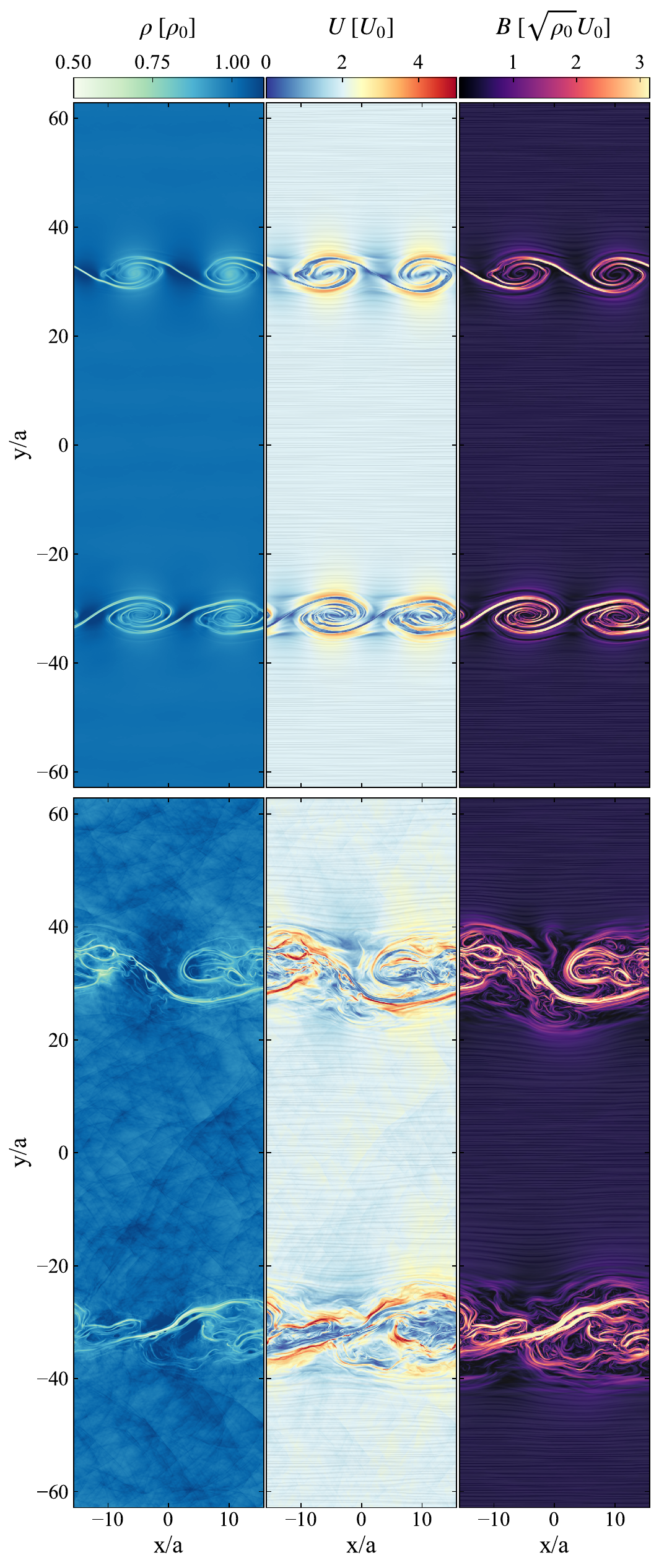}
    \caption{Evolution of the Kelvin-Helmholtz instability showing snapshots at $t = 40$ (top) and $80\,a/U_0$ (bottom). Left column: background fluid density $\rho$. Middle column: velocity field $\boldsymbol{u}$ with streamlines indicating flow direction. Right column: magnetic field $\boldsymbol{B}$ with field lines shown. Colors represent their respective normalized magnitudes.} 
    \label{fig:KHI}
\end{figure}

Figure \ref{fig:KHI} illustrates the typical evolution of shear-driven turbulence in our simulations. During the linear growth phase ($0 < U_0t/a \lesssim 30$), the background flow remains laminar while small-amplitude perturbations grow exponentially (see Appendix \ref{details}). The flow then enters the nonlinear phase at $t \sim 40\,a/U_0$, developing two fully rolled-up vortices along each shear layer. The asymmetric evolution between the two shear layers results from different random perturbations seeded along each layer \citep{henri_nonlinear_2013}. Additionally, the vortex roll-up generates waves that propagate into the bulk flow \citep{karimabadi_coherent_2013}. At $t \sim 50\,a/U_0$, magnetic stresses begin to disrupt the characteristic cat's eye vortices, initiating the transition to fully developed turbulence with a sustained shear profile (i.e., $d\langle u_x \rangle_x / dt \approx 0$). The final state exhibits turbulent mixing layers across both shear regions, rather than large-scale coherent vortices. Appendix~\ref{latetime} presents the same quantities at the much later times $t = 300$ and $600\,a/U_0$, illustrating the persistence of this turbulent state.

\begin{figure}
    \centering
    \includegraphics[width=1.0\linewidth]{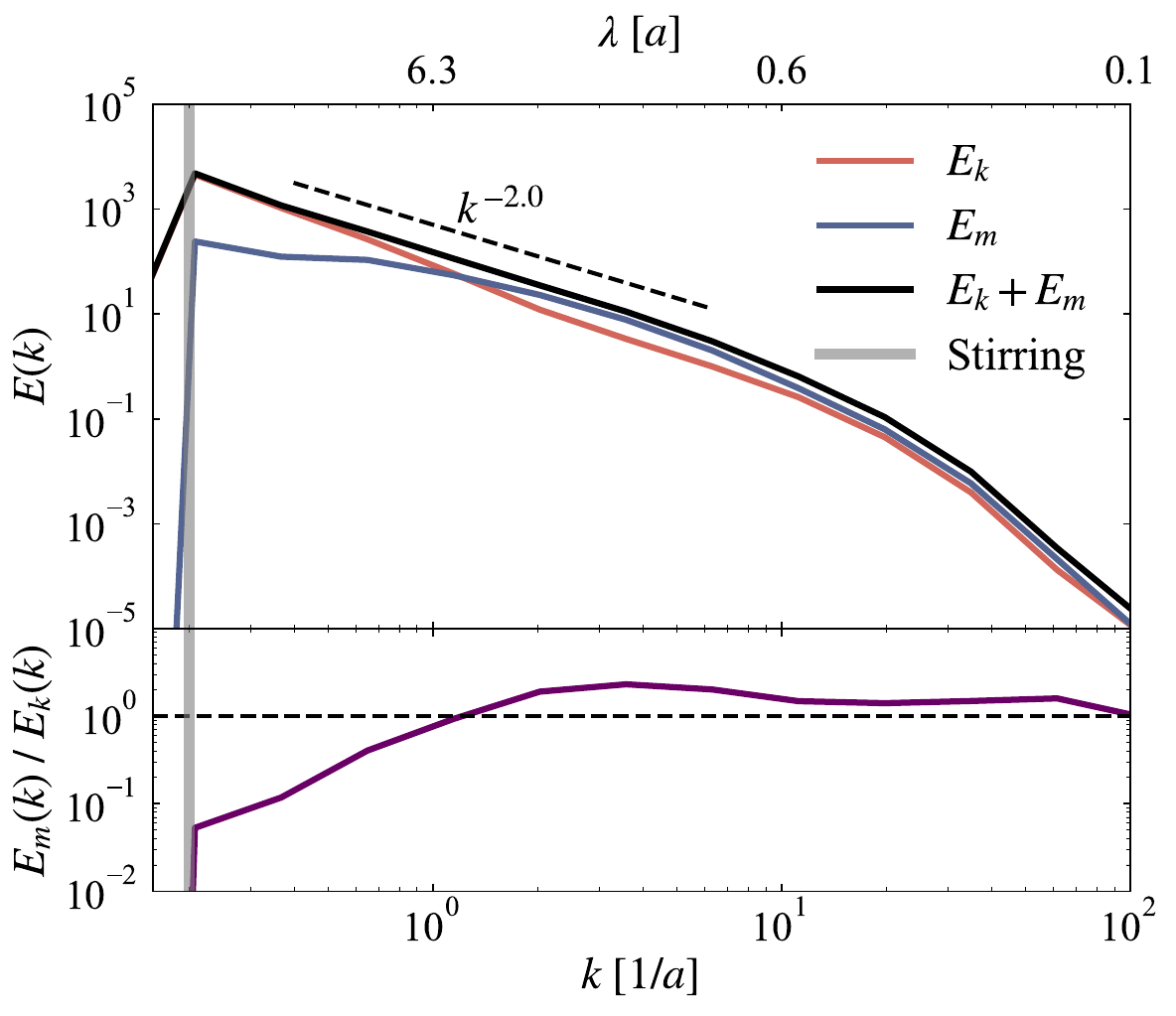}
    \caption{\textbf{Top}: Energy power spectrum of shear-driven turbulence in steady state as a function of wavenumber $k$. Red and blue curves show kinetic ($E_k$) and magnetic ($E_m$) energy densities, respectively; black curve combines both components ($E_k + E_m$). Dashed line indicates $k^{-2}$ scaling in the inertial range. \textbf{Bottom}: Magnetic-to-kinetic energy ratio $E_m/E_k$. Gray vertical line marks the energy injection scale $\lambda_{\rm inj}$ in both panels.} 
    \label{fig:turbulence}
\end{figure}

The stirring force enables the turbulence to reach a quasi-stationary state by $t \sim 150\,a/U_0$, preventing layer broadening \citep{fraser_impact_2021, tripathi_near-cancellation_2022}. In this steady state, energy injection by the stirring force balances the energy dissipation, resulting in a quasi-stationary turbulent energy spectrum in a time-averaged sense, as shown in the top panel of Figure \ref{fig:turbulence}. This figure demonstrates that at large scales, the turbulence is dominated by kinetic energy, particularly the energy injected by the stirring force. At wavelengths larger than the energy injection scale $\lambda_{\rm inj}$, the kinetic energy density exhibits a steep decline due to the finite domain size $L_x$, which constrains the maximum achievable vortex scales.

In comparison, the magnetic energy is not directly coupled to the large-scale stirring mechanism. Instead, the kinetic-to-magnetic energy transfer occurs at smaller scales ($\lambda \lesssim 6\,a$) through vortex stretching and magnetic field line folding \citep{salvesen_quantifying_2014, beattie_energy_2022}. Despite the initially dominant kinetic energy, the magnetic-to-kinetic energy ratio $E_m/E_k$ tends to increase and level off at unity with increasing wavenumber $k$ (see the bottom panel of Figure~\ref{fig:turbulence}). This asymptotic behavior indicates that kinetic and magnetic energies reach approximate equipartition at intermediate-to-large wavenumbers within the inertial range. The out-of-plane field $B_z$ is initially zero and is generated only through the out-of-plane gas motion driven by the particle backreaction force. It saturates at an energy roughly one order of magnitude below that of $B_y$ and two orders of magnitude below that of $B_x$, including the mean field.

\begin{figure}
    \centering
    \includegraphics[width=1.0\linewidth]{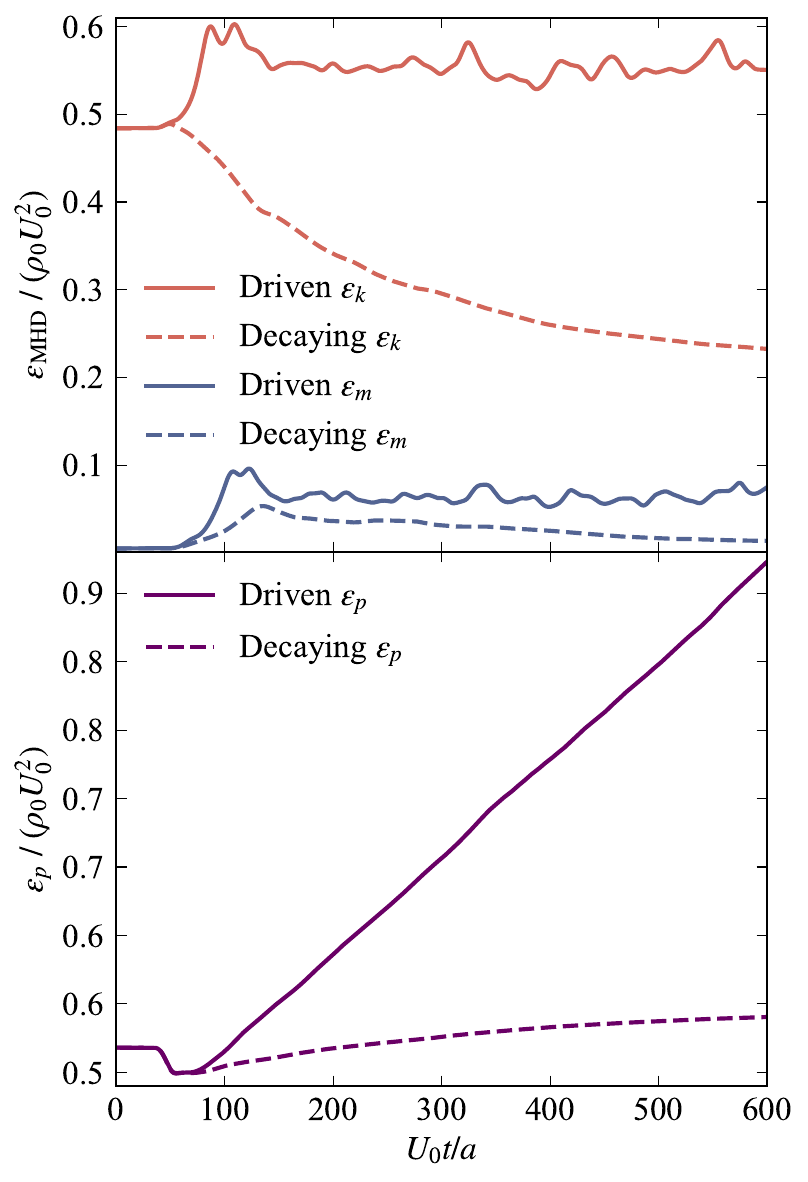}
    \caption{Time evolution of volume-averaged energy densities in driven (solid) and freely decaying (dashed) turbulence. \textbf{Top}: Background fluid kinetic ($\varepsilon_k$, red) and magnetic ($\varepsilon_m$, blue) energy densities vs. time $U_0t/a$. \textbf{Bottom}: Particle energy density $\varepsilon_p$ vs. time.} 
    \label{fig:steady_state}
\end{figure}

The balance between energy injection rate and dissipation rate also sustains kinetic and magnetic energy reservoirs in the steady state, as demonstrated in the top panel of Figure \ref{fig:steady_state}. These abundant turbulent energy reservoirs generate strong motional electric fields $\boldsymbol{E} = -\boldsymbol{u}\times\boldsymbol{B}/c$, which are necessary for particle acceleration. As shown in the bottom panel of Figure \ref{fig:steady_state}, the average particle energy density $\varepsilon_p$ exhibits nearly linear growth with time following the onset of turbulence, indicating sustained particle energization throughout the simulation. This sustained energization persists across a wide range of magnetizations ($3 \leq M_A \leq 40$), from near the KHI stability threshold to the weakly magnetized regime, confirming the generality of the result.

In contrast, freely decaying turbulence fails to sustain particle energization. Without external forcing (i.e., $F_{\text{stir}} = 0$), turbulent energy cascades from large-scale motions to progressively smaller scales, where it dissipates at the grid scale, causing the system to decay rapidly \citep{ryu_magnetohydrodynamic_2000, salvesen_quantifying_2014}. This energy depletion weakens the accelerating electric field over time, resulting in the saturation of particle energy density (see the bottom panel of Figure~\ref{fig:steady_state}). Our results thus demonstrate that sustained particle energization requires driven turbulence with maintained shear profiles and stable energy reservoirs, rather than freely decaying turbulence.

We also examined the role of particle backreaction in sustaining this energization. In an otherwise identical driven simulation evolved without backreaction, corresponding to the test-particle limit, both the volume-averaged kinetic and magnetic energy densities and the turbulent energy spectra agree with the fiducial run to within their temporal fluctuations (not shown). Although $\varepsilon_p$ eventually exceeds $\varepsilon_k$ (Figure~\ref{fig:steady_state}), the growth is powered by the external driving rather than by depleting the turbulent reservoirs, and the sustained energization is therefore not an artifact of the substantial particle energy density adopted in our setup.

\subsection{Particle Energy Distribution} \label{spectrum}
\begin{figure}
    \centering
    \includegraphics[width=1.0\linewidth]{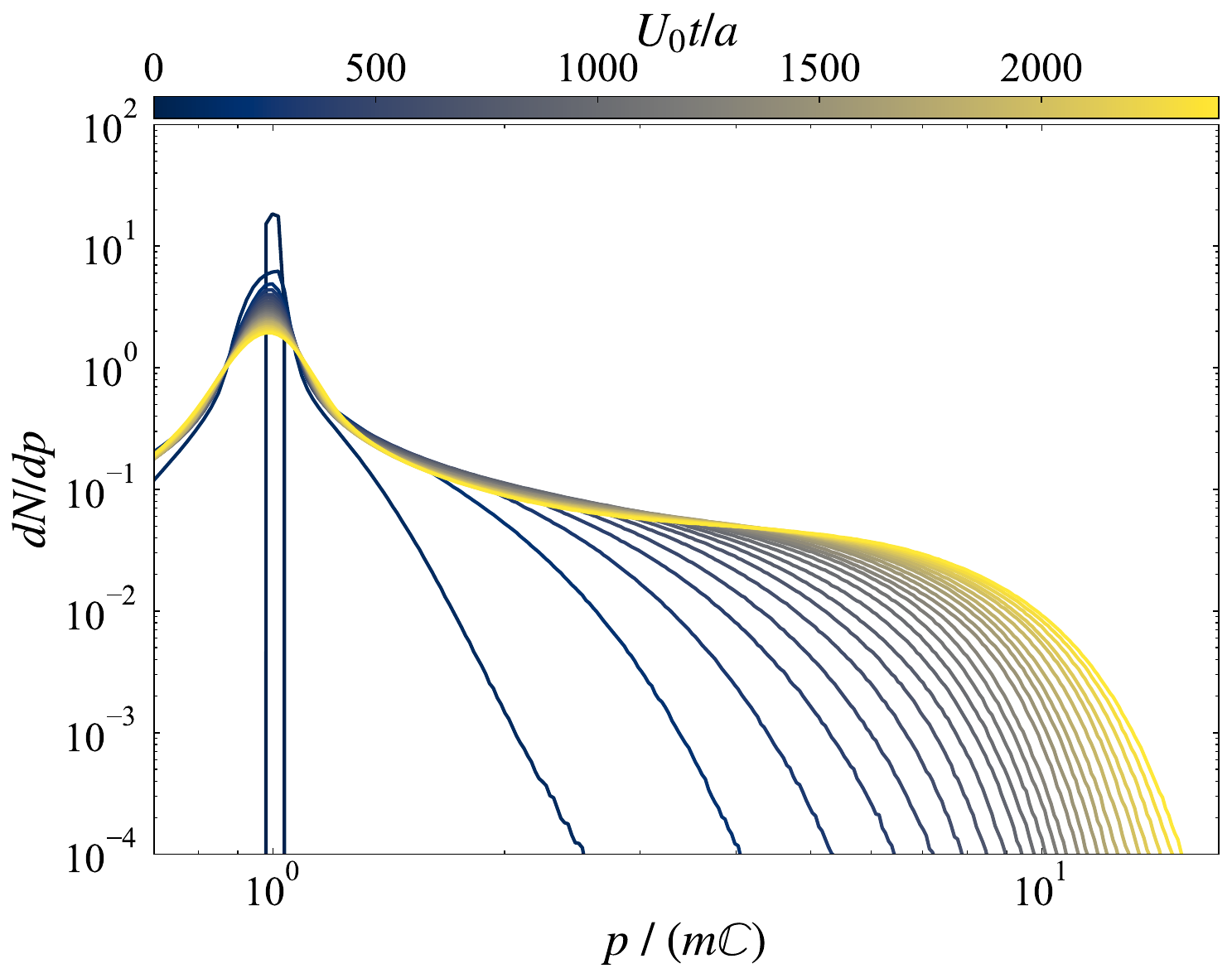}
    \caption{Evolution of the particle momentum distribution $dN/dp$ from $U_0t/a = 0$ to $2400$ (color-coded by time). The initially monoenergetic distribution develops a substantial non-thermal tail, with the high-energy cutoff advancing continuously through sustained acceleration.}
    \label{fig:par_spec}
\end{figure}

While the evolution of average particle energy density (Figure \ref{fig:steady_state}) indicates a sustained heating of the particle population, the energy-dependent distribution of non-thermal particles is of particular interest due to its connections to emission signatures. Therefore, we measured the energy spectrum of all particles in our simulations at different times, shown in Figure \ref{fig:par_spec}.

Initially described by a Dirac delta distribution $f(p, t=0) \approx \delta(p - p_0)$, the particle energy spectrum undergoes significant broadening following the turbulence onset at $t \sim 50\,a/U_0$. While the energy peak remains centered at $p_0$, consisting primarily of particles in the bulk flow that rarely cross shear layers, a substantial non-thermal tail rapidly emerges from particles frequently interacting with turbulent layers. The spectrum is therefore a superposition of two populations rather than a single accelerated distribution: a quasi-static bulk component centered at $p_0$ and an accelerated component that broadens and advances with time. At intermediate momenta, this superposition can be approximately characterized by $f(p) \propto p^{-\alpha}$ with an effective index harder than unity. This does not represent a steady-state power law. In the standard argument for a stationary Fermi spectrum \citep{bell_acceleration_1978}, the power law follows from the combination of the energy gain per cycle and the escape probability per cycle, with continuous injection at low energies replenishing the population. Neither ingredient is present here. Particles cannot escape from the periodic simulation domain, and none are injected, since the particle number is fixed at initialization and the background fluid supplies no fresh particles. The same fixed population therefore drifts and diffuses in momentum space rather than settling into a stationary spectrum. Extrapolated as a stationary power law, $\alpha < 1$ would imply formally divergent particle number and energy at the high-momentum end. No such divergence arises here, because particle number is conserved and the accelerated component follows the number-conserving log-normal distribution of Section \ref{acceleration} (Figures \ref{fig:GBM_evo} and \ref{fig:GBM_pdf}), whose moments remain finite at all times. The effective spectral index $\alpha$ decreases monotonically with time, indicating that the non-thermal tail becomes progressively flatter as particles continue to be accelerated to higher energies; Section \ref{acceleration} traces this hardening to the broadening and advance of the log-normal component.

The advance of the high-energy cutoff is initially rapid and decelerates before settling to a near-constant rate, $dp_{\text{max}}/dt \approx \text{const}$, at late times. This late-time behavior is consistent with the stochastic-acceleration timescale $t_{\rm acc} \simeq p^2/D_p$ \citep{schlickeiser_cosmic-ray_1989} evaluated with the measured diffusion coefficient: the implied advance rate, $dp_{\max}/dt \sim p_{\max}/t_{\rm acc} = D_p/p_{\max}$, varies only mildly across the traversed momentum range, because $D_p$ rises through trans-relativistic momenta and turns over near $p/(m\mathbb{C}) \sim 8$ (Figure~\ref{fig:diffusion}; Appendix~\ref{FP}). The same turnover bounds this regime: beyond the peak of $D_p$, the acceleration timescale grows steeply, $t_{\rm acc} \propto p^3$, and becomes comparable to the simulation duration, so the progressive flattening of the spectrum is a transient of the simulated interval rather than a sustainable state, and the cutoff advance is expected to slow further toward $p_{\max} \propto t^{1/3}$ in the declining branch. Within the flat portion of the spectrum, which marks the momenta occupied by the accelerated log-normal component (Section~\ref{acceleration}), absolute momentum gains are largest where $D_p$ peaks, at $p/(m\mathbb{C}) \sim 7$.

\begin{figure}
    \centering
    \includegraphics[width=1.0\linewidth]{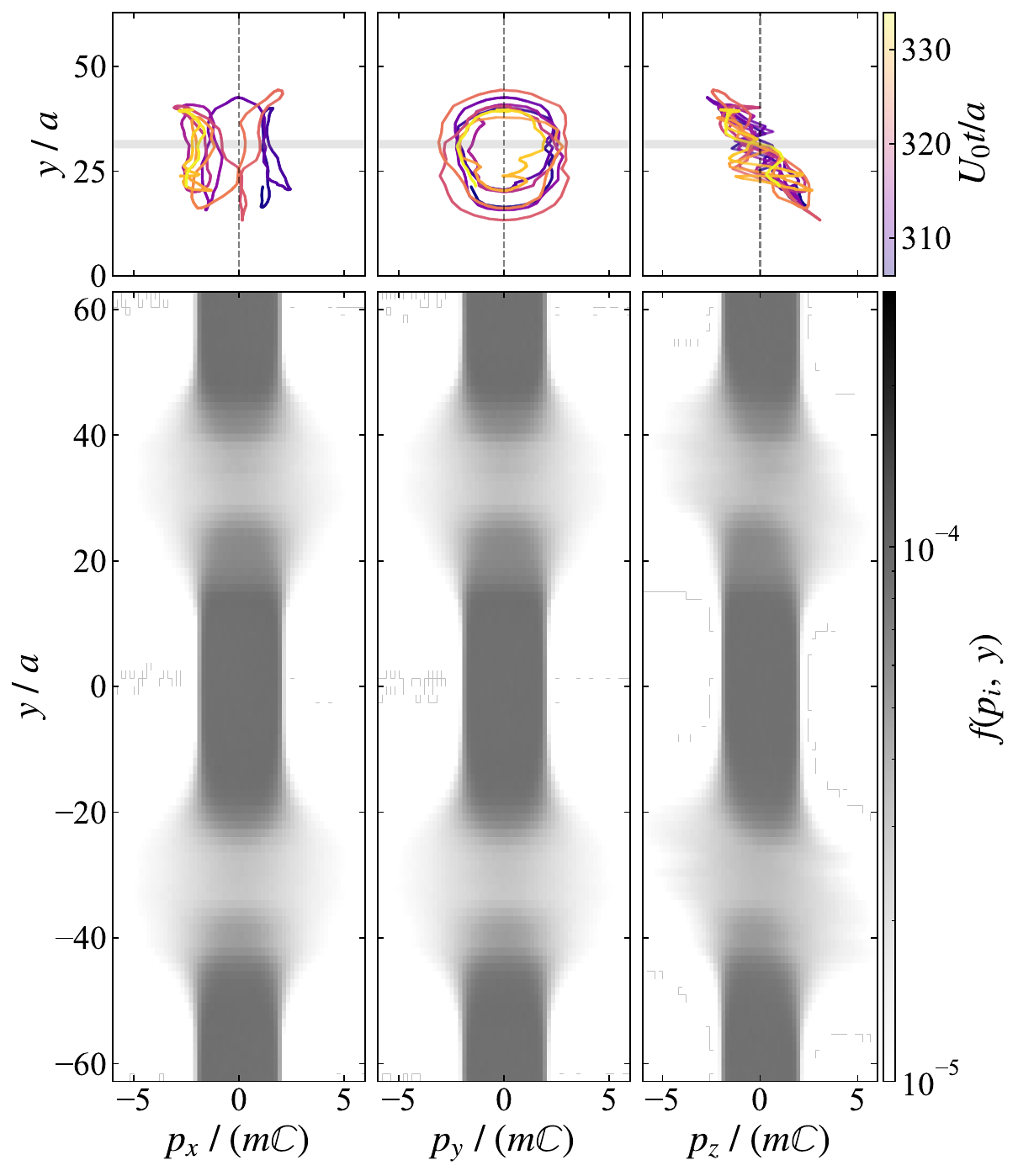}
    \caption{Phase-space evolution of particles in shear-driven turbulence. \textbf{Top:} Trajectory of a single particle in $(p_x, y)$, $(p_y, y)$, and $(p_z, y)$ planes, color-coded by time from $t = 310\,a/U_0$ to $330\,a/U_0$. Dashed lines indicate zero momentum; gray bands mark shear layer locations at $y = y_{1,2}$. \textbf{Bottom:} Phase-space distributions $f(p_i/m, y)$ for the particle ensemble at $t = 1200\,a/U_0$, showing momentum diffusion at shear layers and the antisymmetric structure in $p_z$.} 
    \label{fig:phase_space}
\end{figure}

During this non-thermal acceleration, particles undergo momentum diffusion in phase space, as shown in Figure~\ref{fig:phase_space}, revealing the underlying particle dynamics in the turbulence. The top row shows the phase-space trajectory of a single particle in the $(p_x, y)$, $(p_y, y)$, and $(p_z, y)$ planes as functions of time. Since the stirring force maintains the shear profile, turbulence remains localized near the shear layers at $y = y_{1,2}$, while the initial uniform magnetic field $\boldsymbol{B}_0 = B_0\,\hat{\boldsymbol{x}}$ persists in the bulk flow regions. As a result, particles gyrate primarily in the $y$-$z$ plane perpendicular to $\boldsymbol{B}_0$ and repeatedly cross the shear layers. This gyromotion produces a linear correlation in the $(p_z, y)$ plane with slope $dp_z/dy = -qB_0/c$ and circular patterns in the $(p_y, y)$ plane, consistent with cyclotron motion around the $x$-directed magnetic field. Meanwhile, the $x$-component of momentum $p_x$ varies randomly, exhibiting no correlation with position $y$, indicating stochastic scattering independent of the background shear flow.

The bottom row of Figure~\ref{fig:phase_space} presents the phase-space distribution $f(p_i, y)$ for the entire particle ensemble at $t = 1200\,a/U_0$. The momentum variance $\langle \Delta p_i^2 \rangle$ for all three components ($i = x, y, z$) peaks at the shear layer locations $y = y_{1,2}$ and reaches minima in the bulk flow regions, confirming that particle acceleration occurs primarily through shear layer interactions. The distributions of $p_x$ and $p_y$ broaden symmetrically about their initial values, maintaining $\langle p_x \rangle = \langle p_y \rangle = 0$. In contrast, the $p_z$ distribution exhibits antisymmetric structure: particles obtain negative $p_z$ for $y > y_{1,2}$ and positive $p_z$ for $y < y_{1,2}$. This antisymmetric pattern, with $\partial \langle p_z \rangle / \partial y < 0$, arises from the coherent gyrophase organization of particles crossing the shear layers.

\subsection{Acceleration Mechanism} \label{acceleration}
\begin{figure}
    \centering
    \includegraphics[width=1.0\linewidth]{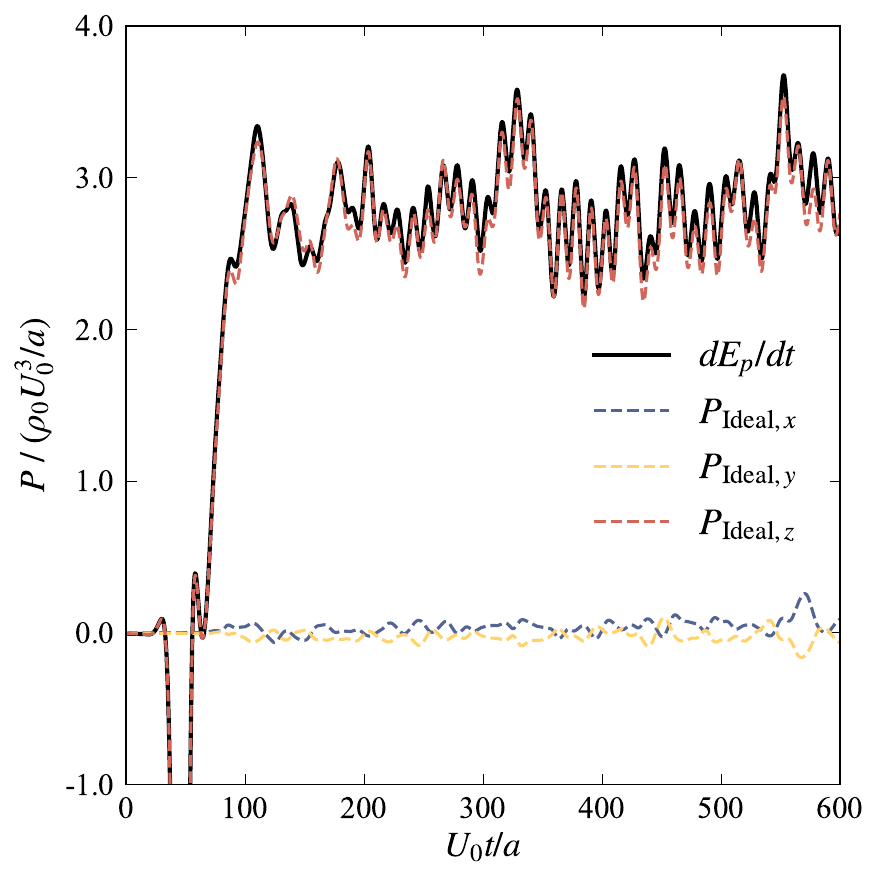}
    \caption{Time evolution of the power delivered to the particles, decomposed into $x$, $y$, and $z$ components of the electric field. The total power, defined as the time-derivative of total particle energy $dE_p / dt$, is shown in black solid line. Three components of the power delivered by the motional electric field ($P_{\rm{ideal},x}$, $P_{\rm{ideal},y}$, and $P_{\rm{ideal},z}$) are represented by dashed lines.} 
    \label{fig:power}
\end{figure}

To quantitatively study particle acceleration in shear-driven turbulence, we decompose the power delivered by the ideal motional electric field into Cartesian components: $P_{\rm{ideal},i} = q E_i v_i$ for $i \in \{x,y,z\}$ (Figure \ref{fig:power}). The total particle energy gain rate $dE_p/dt$ initially increases proportionally with the turbulent kinetic energy growth, reaching a quasi-steady value of $\langle dE_p/dt \rangle \approx 3\,\rho_0 U_0^3/a$ once turbulence saturates at $t \gtrsim 100\,a/U_0$. The z-component of the motional electric field, $E_z = -(u_xB_y - u_yB_x)/c$, dominates the acceleration process, contributing $\gtrsim 98\%$ of the total power throughout the nonlinear phase. In contrast, the in-plane components $P_{\rm{ideal},x}$ and $P_{\rm{ideal},y}$ oscillate around zero with amplitudes $\lesssim 0.02\langle P_{\rm{ideal},z}\rangle$. This strong anisotropy demonstrates that particles preferentially gain energy through interactions with out-of-plane electric fields generated by the cross product of in-plane velocity field and magnetic field fluctuations.

\begin{figure}
    \centering
    \includegraphics[width=1.0\linewidth]{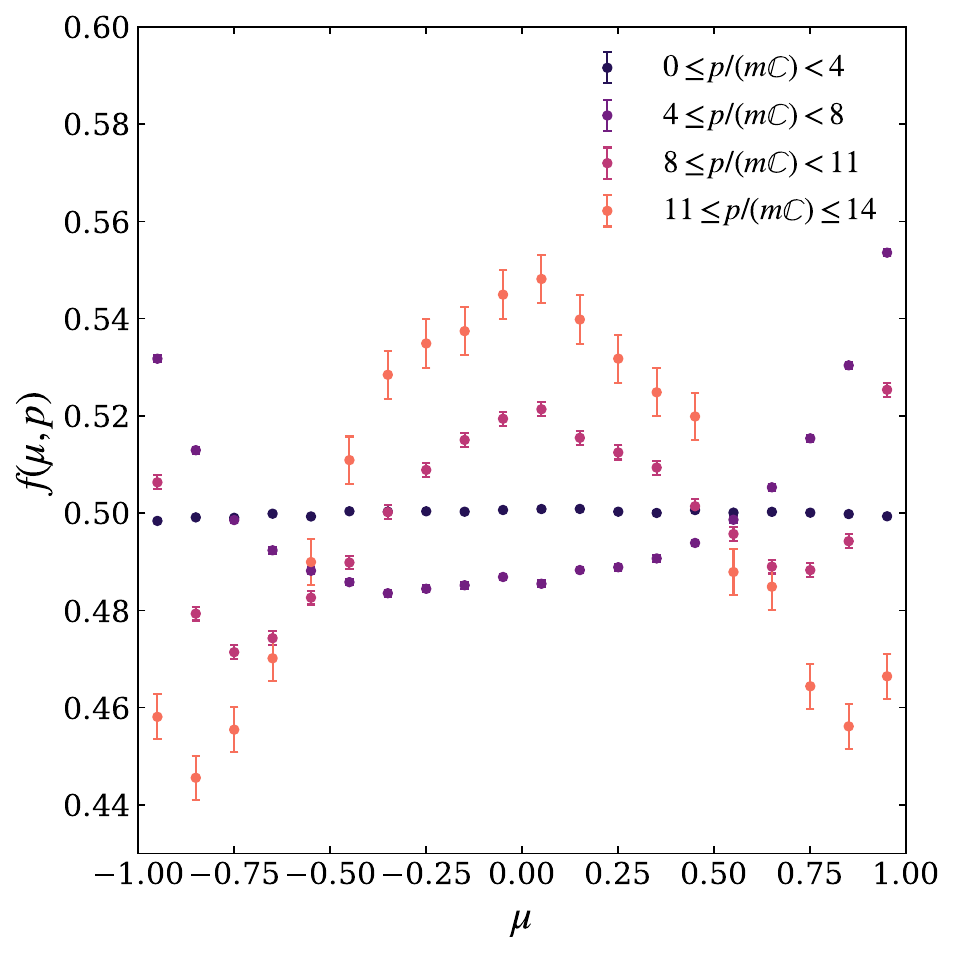}
    \caption{Pitch angle distribution $f(\mu, p)$ at $t = 2400\,a/U_0$, where $\mu$ is the pitch angle cosine, with particle velocity and magnetic field both evaluated in the local $\boldsymbol{E}\times\boldsymbol{B}$ frame. Particles are binned by momentum, and error bars show the $1\sigma$ statistical uncertainty of each bin. Particles with $p/(m\mathbb{C}) < 4$ remain isotropic, while at intermediate momenta the distribution is enhanced near $|\mu| = 1$. Particles with $p/(m\mathbb{C}) \gtrsim 8$ accumulate near $\mu \approx 0$ as their gyroradii exceed the turbulent layer width.}
    \label{fig:pitch_dist}
\end{figure}

The dominance of $E_z$-driven acceleration naturally produces pitch angle anisotropy once particle gyroradii exceed the turbulent layer width (Figure \ref{fig:pitch_dist}). We measure the pitch angle cosine $\mu \equiv \boldsymbol{v} \cdot \boldsymbol{B}/ (|\boldsymbol{v}||\boldsymbol{B}| ) \in [-1,1]$, with $\boldsymbol{v}$ and $\boldsymbol{B}$ both evaluated in the local $\boldsymbol{E}\times\boldsymbol{B}$ frame in which the motional electric field vanishes, and we bin particles by momentum to construct the pitch angle distribution $f(\mu, p)$, normalized to $\int_{-1}^{1} f(\mu, p)\, d\mu = 1$ within each momentum bin, so that an isotropic distribution corresponds to $f = 1/2$. Figure \ref{fig:pitch_dist} shows $f(\mu, p)$ at $t = 2400\,a/U_0$, the final epoch of Figure \ref{fig:par_spec}; the distributions evolve together with the spectrum, each momentum bin developing its anisotropy as accelerated particles populate it, while the lowest bin remains isotropic throughout the simulation. For low-energy particles with $p/(m\mathbb{C}) \lesssim 4$, having small gyroradii and thus confined within the turbulence, frequent turbulent scattering maintains equal probability across $\mu$. However, sustained acceleration by $E_z$ systematically increases $p_\perp = p\sqrt{1-\mu^2}$, causing gyroradii to grow as $r_{\rm c} \propto p_\perp$. Once $p/(m\mathbb{C}) \gtrsim 8$, particles increasingly sample the laminar regions where $\delta B/B_0 \ll 1$, and a reduced scattering rate preserves their enhanced perpendicular momentum. As a result, particles begin to concentrate near $\mu = 0$, with amplitudes well above the $1\sigma$ statistical uncertainties. This mechanism differs fundamentally from the mirror acceleration in relativistic turbulence reported by \citet{das_studying_2025}, where anisotropy arises from the electric field $E_{\perp}$ induced by temporal changes of magnetic flux enclosed by gyro-orbits. Here, anisotropy emerges from the interplay between perpendicular electric field acceleration ($E_z \perp B_x$) and the spatial confinement of turbulent scattering, with particles retaining enhanced $p_\perp$ once they gyrate out of the turbulence.

\begin{figure}
    \centering
    \includegraphics[width=1.0\linewidth]{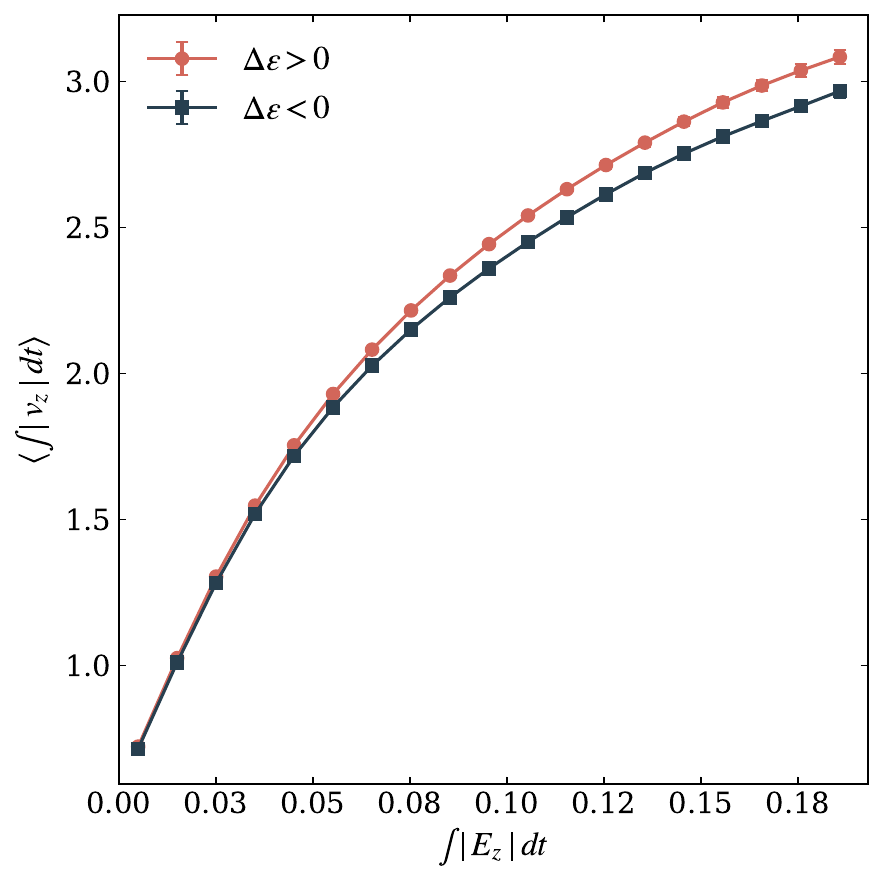}
    \caption{Asymmetry in particle path length during electric field interactions. The plot shows the average distance traveled along the $z$-direction, $\langle \int |v_z| dt\rangle$, as a function of the integrated electric field magnitude $\int |E_z| dt$ for individual aligned (red circles) and anti-aligned (blue squares) trajectory segments. Particles systematically travel longer distances during accelerating phases ($\Delta\epsilon > 0$) than during decelerating phases ($\Delta\epsilon < 0$) due to the velocity perturbation induced by the field. Error bars indicate standard errors.}
    \label{fig:distortion}
\end{figure}

Particles gyrating through turbulent layers experience stochastic energy changes $\Delta\epsilon$ governed by the instantaneous alignment of $\boldsymbol{v}$ and $\boldsymbol{E}$, which reverses sign multiple times along the gyro-orbit. While individual interactions may result in energy gain or loss, the cumulative effect is a net energization. This net gain stems from the systematic distortion of the gyro-orbit by the electric field. To illustrate this process, we decompose particle trajectories into many segments, where $\boldsymbol{v}$ is either aligned or anti-aligned with the local $\boldsymbol{E}$ for a duration much shorter than the gyroperiod. In aligned segments ($\boldsymbol{v} \cdot \boldsymbol{E} > 0$), the particle increases speed relative to its unperturbed gyro-orbit. This velocity perturbation $\delta \boldsymbol{v}$ introduces an \textit{additional} spatial displacement $\delta z$ along the electric field $\boldsymbol{E}$ that predominantly points in the $z$-direction. This modification extends the spatial path length over which the accelerating force acts. Conversely, in anti-aligned segments ($\boldsymbol{v} \cdot \boldsymbol{E} < 0$), the particle loses speed relative to the unperturbed orbit, which effectively shortens the interaction path. Although our simulation is two-dimensional, the three-dimensional particle velocity evolution (Equations \ref{eq:eom_start}--\ref{eq:eom_end}) allows us to infer the effective path length change in the z-direction from the measured velocity changes. Figure \ref{fig:distortion} confirms this asymmetry: particles consistently travel larger distances during accelerating phases than during decelerating phases for a given electric field strength.

\begin{figure}
    \centering
    \includegraphics[width=1.0\linewidth]{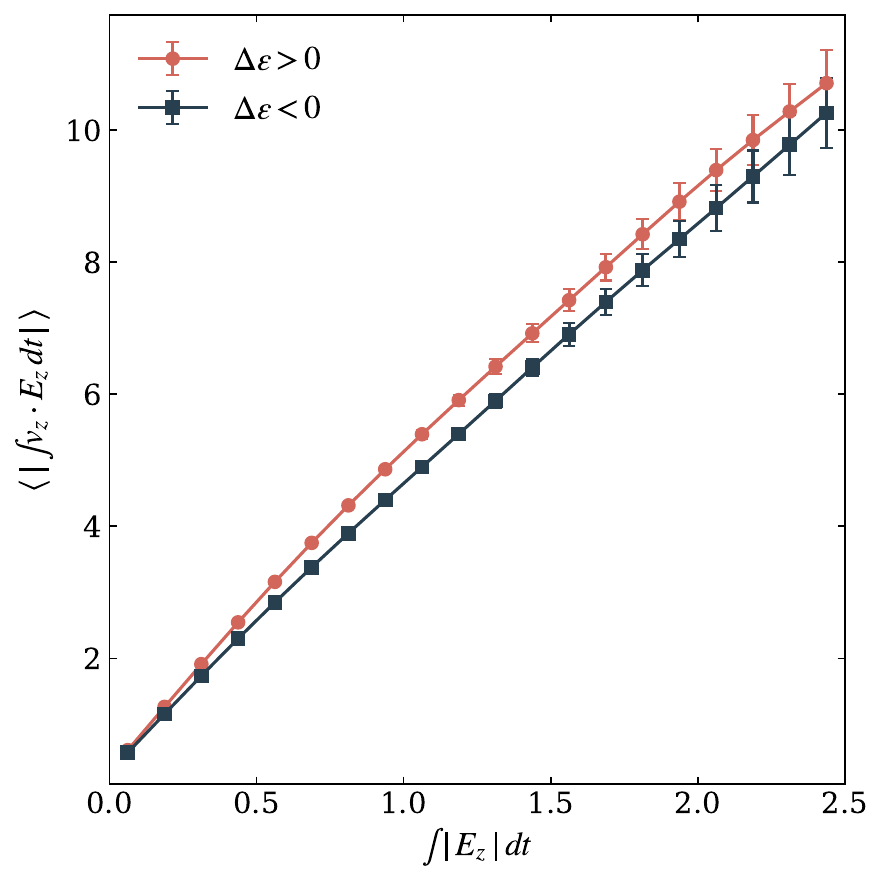}
    \caption{Net energy gain resulting from the path length asymmetry. The plot shows the average magnitude of energy change, $\langle|\int v_z E_z dt|\rangle$, versus the integrated field magnitude over individual gyrocycles. Red circles represent net energy gains ($\Delta\epsilon > 0$), while blue squares represent net losses ($\Delta\epsilon < 0$). Energy gains consistently exceed losses for a given field strength, confirming that the orbit distortion leads to systematic particle energization. Error bars indicate standard errors.}
    \label{fig:gain_loss}
\end{figure}

Since the work done on a particle corresponds to the path integral of the force ($\Delta \epsilon \approx q \int E_z dz$), the energy gained over the extended path in aligned segments exceeds the energy lost over the shortened path in anti-aligned segments. Figure \ref{fig:gain_loss} validates this mechanism: when averaged over individual gyrocycles, net energy gains ($\Delta\epsilon > 0$) systematically exceed losses ($\Delta\epsilon < 0$) whenever particles sample non-zero electric fields. Because the orbit distortion is small compared to the unperturbed gyromotion, the energy evolution follows a biased random walk. The dominant stochastic fluctuations, driven by the spatiotemporal variability of the turbulent electric field, cause energy diffusion, while the small positive drift per encounter leads, over many encounters, to systematic acceleration---characteristic of a second-order Fermi process.

To quantify this acceleration mechanism, consider a trajectory segment of duration $\tau$, referred to here as the correlation time, over which the particle velocity remains either aligned or anti-aligned with the electric field. In this short interval, we assume the electric field $E_z$ is constant and the rotation of the particle velocity vector by the magnetic field is negligible. The electric field induces a velocity perturbation $\delta v_z$ in the direction of the force:
\begin{equation}
    \delta v_z \approx \frac{q\tau}{\gamma m} E_z \left( 1 - \frac{v_z^2}{\mathbb{C}^2} \right),
\end{equation}
where the relativistic factor $1/(\gamma m)$ accounts for the increased particle inertia at high velocities. The work done along the unperturbed path, $W_0 = q E_z v_z \tau$, averages to zero due to the random alignment of $\boldsymbol{v}$ and $\boldsymbol{E}$, but its fluctuations dominate the energy diffusion. The perturbation $\delta v_z$ further results in a second-order energy change per segment:
\begin{equation}\label{delta_eps}
    \delta \epsilon \approx \frac{q^2 E_z^2 \tau^2}{2 \gamma m} \left( 1 - \frac{v_z^2}{\mathbb{C}^2} \right).
\end{equation}
Since $v_z^2 < \mathbb{C}^2$, this term is strictly positive regardless of the instantaneous alignment, leading to a net energy gain over many interactions. For a detailed derivation, please see Appendix \ref{derivation}.

In the context of shear layer crossing, the correlation timescale is determined by the layer width and the perpendicular velocity, $\tau \sim a/|v_y|$, approximating the electric field as constant over the shear layer half-width $a$. Assuming a gyrotropic distribution (Figure \ref{fig:phase_space}), the mean squared and mean energy changes per crossing, governed respectively by the linear work $W_0$ and by the distortion term $\delta\epsilon$ (Equation \ref{delta_eps}), scale as:
\begin{equation}\label{mean_square}
    \left\langle (\Delta \epsilon)^2 \right\rangle \sim \left\langle \epsilon_*^2 \frac{v_z^2}{v_y^2} \right\rangle \sim \epsilon_*^2,
\end{equation}
\begin{equation}\label{mean_change}
    \left\langle \Delta \epsilon \right\rangle \sim \frac{\epsilon_*^2}{\epsilon},
\end{equation}
where $\epsilon_* \equiv qE_za$ represents the characteristic energy change per crossing. The energy diffusion $\langle (\Delta \epsilon)^2 \rangle$ depends primarily on the electric field strength and is independent of particle energy for high-energy particles.

\begin{figure}
    \centering
    \includegraphics[width=1.0\linewidth]{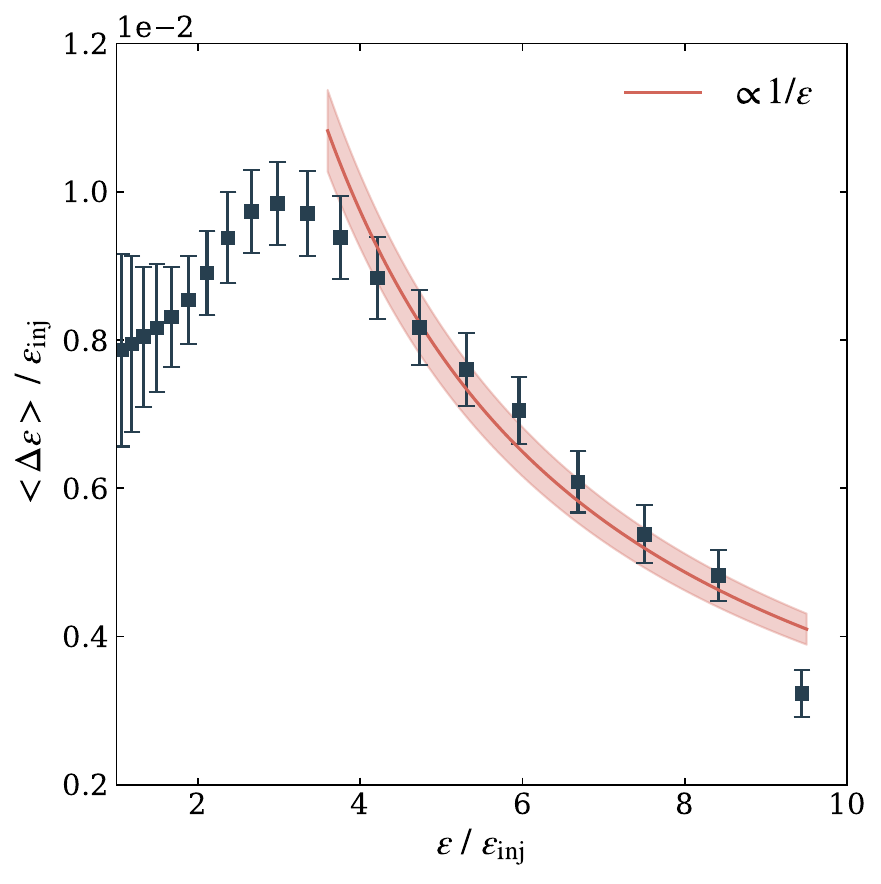}
    \caption{Particle mean energy change $\langle\Delta\epsilon\rangle/\epsilon_{\text{inj}}$ as a function of energy $\epsilon/\epsilon_{\text{inj}}$. Squares show simulation data; the solid line shows $\propto 1/\epsilon$ scaling, with the shaded region representing a 95\% confidence interval. At high energies, the inverse scaling arises from relativistic inertia, which suppresses the velocity distortion $\delta v$ required for net energy gain.}
    \label{fig:mean_energy_change}
\end{figure}

\begin{figure}
    \centering
    \includegraphics[width=1.0\linewidth]{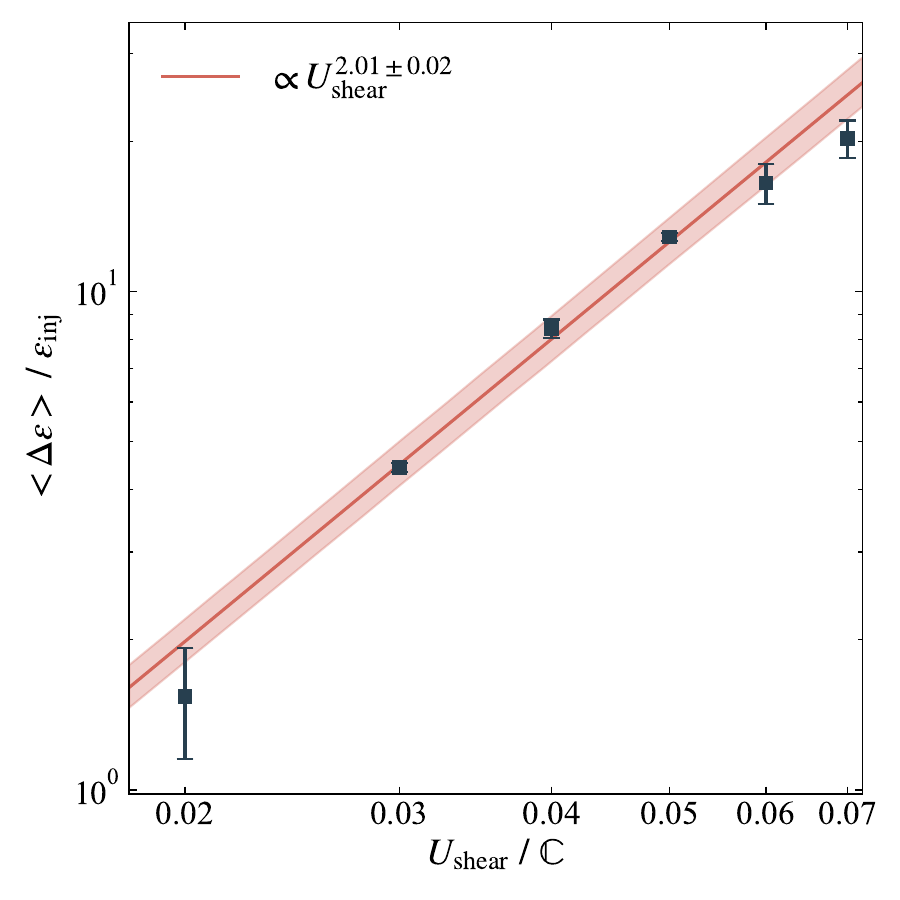}
    \caption{Particle mean energy change $\langle\Delta\epsilon\rangle/\epsilon_{\text{inj}}$ as a function of shear velocity. The solid line represents the linear regression fit, with the shaded region representing a 95\% confidence interval. The best-fit slope $b = 2.01 \pm 0.02$ confirms the quadratic scaling $\langle \Delta \epsilon \rangle \propto U_{\rm{shear}}^2$ characteristic of second-order Fermi acceleration.} 
    \label{fig:second_order}
\end{figure}

Equation \ref{mean_change} implies that the mean energy gain per crossing scales as $\langle \Delta \epsilon \rangle \propto 1/\epsilon$ for a fixed field strength (Figure \ref{fig:mean_energy_change}). In the low-energy regime ($\epsilon / \epsilon_{\rm{inj}} \lesssim 3$), $\langle \Delta \epsilon \rangle$ initially rises as growing gyroradii facilitate shear crossing, extending the effective interaction path for particles originating outside the layers. Once this path saturates at $\sim$$\Delta = 2\,a$, the inverse scaling $\langle \Delta \epsilon \rangle \propto 1/\epsilon$ dominates due to relativistic inertia. Specifically, as the effective mass $\gamma m$ grows, the electric force induces a smaller velocity distortion $\delta v \propto (\gamma m)^{-1}$, thereby reducing the energy gain per crossing. This inverse scaling is also the behavior expected for non-resonant interaction with effectively uncorrelated field structures, the relevant regime once particle gyroradii exceed the coherence scale of the fields \citep{demidem_particle_2020}. The momentum impulse per crossing, $\delta p \approx q E_z \tau$, is then independent of particle energy, and the mean gain, second order in this impulse, decreases as $1/\epsilon$ (Appendix~\ref{derivation}). Furthermore, because suprathermal particles ($\epsilon / \epsilon_{\rm{inj}} \gtrsim 4$) operate at scales where magnetic field amplification decouples from the shear velocity (Figure \ref{fig:turbulence}), the characteristic energy change follows $\epsilon_*^2 \propto E_z^2 \propto U_{\rm{shear}}^2$. The quadratic scaling, verified in Figure \ref{fig:second_order}, suggests a second-order acceleration with a proportionality coefficient of order unity. This microscopic energization leads to significant macroscopic energy conversion, with particles absorbing approximately 10\% of the total injected power (see Section \ref{discussion}), confirming the efficiency of the acceleration mechanism.

To connect microscopic scattering to the macroscopic distribution, we model the stochastic energy evolution as a geometric Brownian motion (GBM), where the noise amplitude scales with particle momentum. Multiplicative stochastic processes of this form arise widely in physics \citep{redner_random_1990}, a classical example being the log-normal statistics of turbulent energy dissipation \citep{kolmogorov_refinement_1962}. The particle momentum evolution follows the stochastic differential equation (SDE) in It\^o form:
\begin{equation}\label{SDE}
    dp_t = \xi p_t dt + \sigma p_t dW_t,
\end{equation}
where $p_t$ is the time-varying particle momentum, $\xi$ is the drift coefficient representing systematic energy gain, and $\sigma$ is the noise coefficient capturing stochastic fluctuations. This SDE directly relates to the Fokker-Planck equation with momentum-dependent diffusion coefficient $D_p \propto p^2$ (see Appendix \ref{FP} for details). Here, $dW_t$ is the non-biased Brownian motion with $\langle dW_t \rangle = 0$, arising from the random alignment between $\boldsymbol{v}$ and $\boldsymbol{E}$. The stochastic term scales as
\begin{equation}
    dW_t \propto \cos\theta_t = \frac{\boldsymbol{v} \cdot \boldsymbol{E}}{|\boldsymbol{v}||\boldsymbol{E}|},
\end{equation}
where $\theta_t$ is the instantaneous angle between velocity and electric field vectors. While Equations~\ref{mean_square} and \ref{mean_change} describe energy-independent kicks per crossing at fixed field strength, the GBM form serves here as a compact empirical description of the frequently crossing population over the simulated dynamic range, with $\xi$ and $\sigma$ treated as fitted coefficients.

The solution to Equation \ref{SDE} yields the expected value and probability distribution:
\begin{equation}\label{expected}
    \mathbb{E}[p_t] = p_0 e^{\xi t},
\end{equation}
\begin{equation}\label{log_normal}
    f(p_t) = \frac{1}{p \sigma \sqrt{2\pi t}} \exp\left( -\frac{\left[\ln(p/p_0) - \left( \xi - \frac{1}{2}\sigma^2 \right)t \right]^2}{2\sigma^2 t} \right),
\end{equation}
where $p_0$ is the initial particle momentum. Equation \ref{log_normal} represents a log-normal distribution, arising from the multiplication of random variables \citep{kahn_note_1973, esmen_lognormality_1977, ott_physical_1990, andersson_mechanisms_2021}. This distribution shape follows from the Central Limit Theorem \citep{feller1991introduction} applied to the logarithm of the momentum: the explicit solution of Equation \ref{SDE}, $p_t = p_0 \exp\left[\left(\xi - \sigma^{2}/2\right)t + \sigma W_t\right]$, shows that $\ln p_t$ accumulates the stochastic increments $dW_t \propto \cos\theta_t$ contributed by successive trajectory segments. Because the alignment between $\boldsymbol{v}$ and $\boldsymbol{E}$ is randomized from one segment to the next, these increments are independent or weakly correlated, and their sum approaches a normal distribution regardless of the distribution of any individual increment. Therefore $\ln p_t$ is normally distributed, and $p_t$ follows the log-normal form of Equation \ref{log_normal}.

\begin{figure*}
    \centering
    \includegraphics[width=1.0\linewidth]{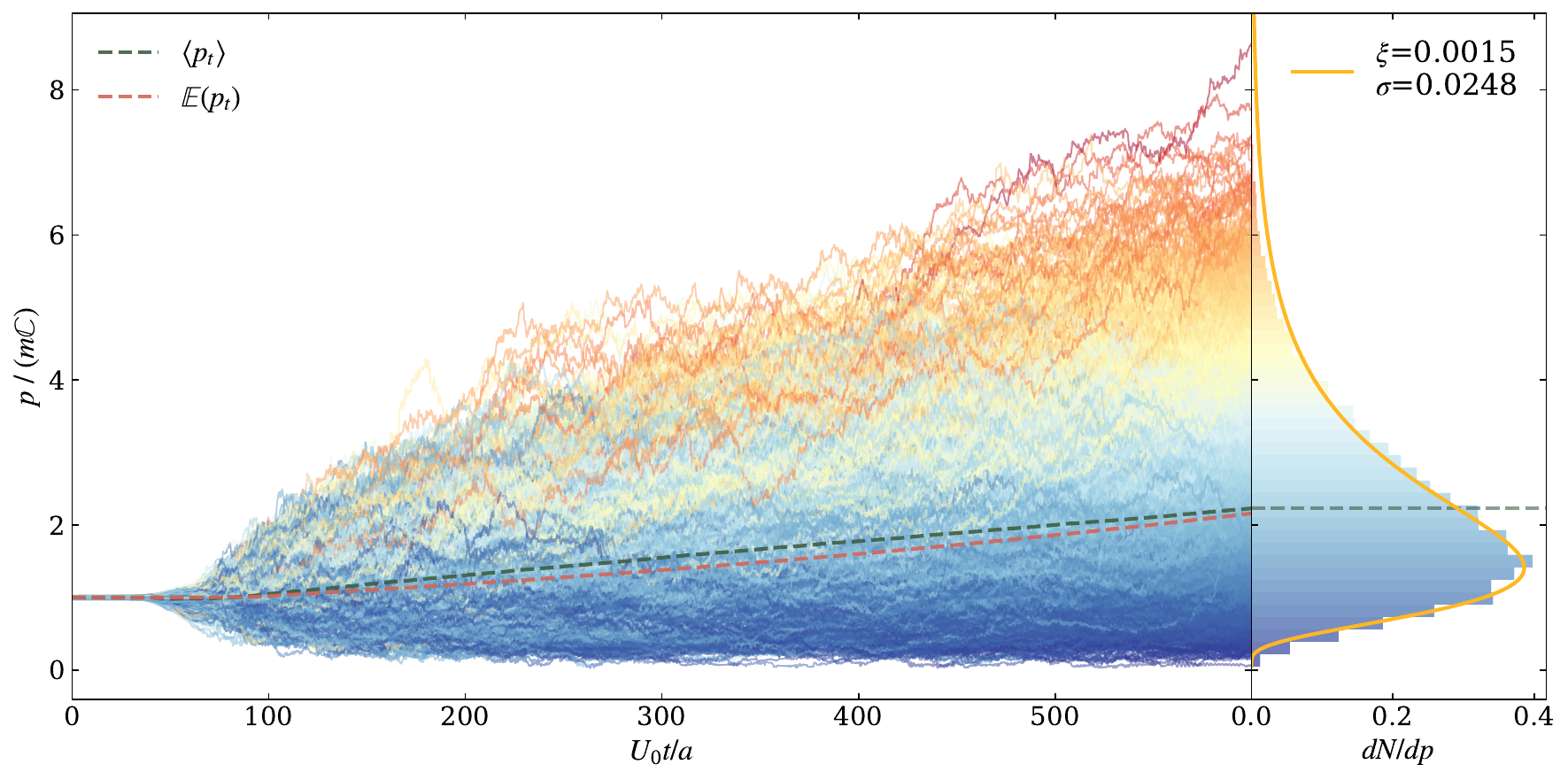}
    \caption{GBM analysis of particles with $N_{\text{cross}} > 40$. Left: Individual particle momentum trajectories color-coded by final momentum, overlaid with the observed mean $\langle p_t\rangle$ (green dashed) and expected value $\mathbb{E}[p_t]$ (red dashed) from Equation \ref{expected} using fitted parameters $\xi = 0.0015$, $\sigma = 0.0248$. Right: Final momentum distribution $dN/dp$ with log-normal fit from Equation \ref{log_normal}.}
    \label{fig:GBM_evo}
\end{figure*}

Particles near shear layers cross the turbulent layers repeatedly, experiencing significantly higher diffusion and acceleration than bulk-flow particles. Figure \ref{fig:GBM_evo} verifies GBM predictions for this frequently crossing population. In the left panel, the observed average momentum $\langle p_t \rangle$ closely follows the expected value $\mathbb{E}[p_t] = p_0 e^{\xi t}$. In the right panel, the final momentum distribution matches the predicted log-normal form (Equation \ref{log_normal}), with best-fit $\xi = 0.0015\,U_0/a$ and $\sigma = 0.0248\,(U_0/a)^{1/2}$. Since $\xi$ and $\sigma$ carry different dimensions, the relevant comparison is the dimensionless ratio $\xi/\sigma^2 \approx 2.4$, which shows that the systematic drift and the diffusion are of comparable strength. This is consistent with Equations \ref{mean_square} and \ref{mean_change}: per encounter, the energy change is dominated by fluctuations, with a second-order bias smaller by a factor of order $\epsilon_*/\epsilon$, but the bias accumulates linearly with the number of encounters $N$ while the random component grows only as $\sqrt{N}$, so over the many crossings that this population undergoes the systematic gain becomes competitive with the diffusive spread. In the log-normal solution, the ratio of the displacement of $\ln p$ to its spread, $(\xi - \sigma^2/2)\sqrt{t}/\sigma$, grows as $\sqrt{t}$ and reaches $\approx 1.2$ at $t = 600\,a/U_0$. The process is thus a momentum diffusion whose drift is generated by the fluctuations themselves rather than by an additional deterministic acceleration.

\begin{figure}
    \centering
    \includegraphics[width=1.0\linewidth]{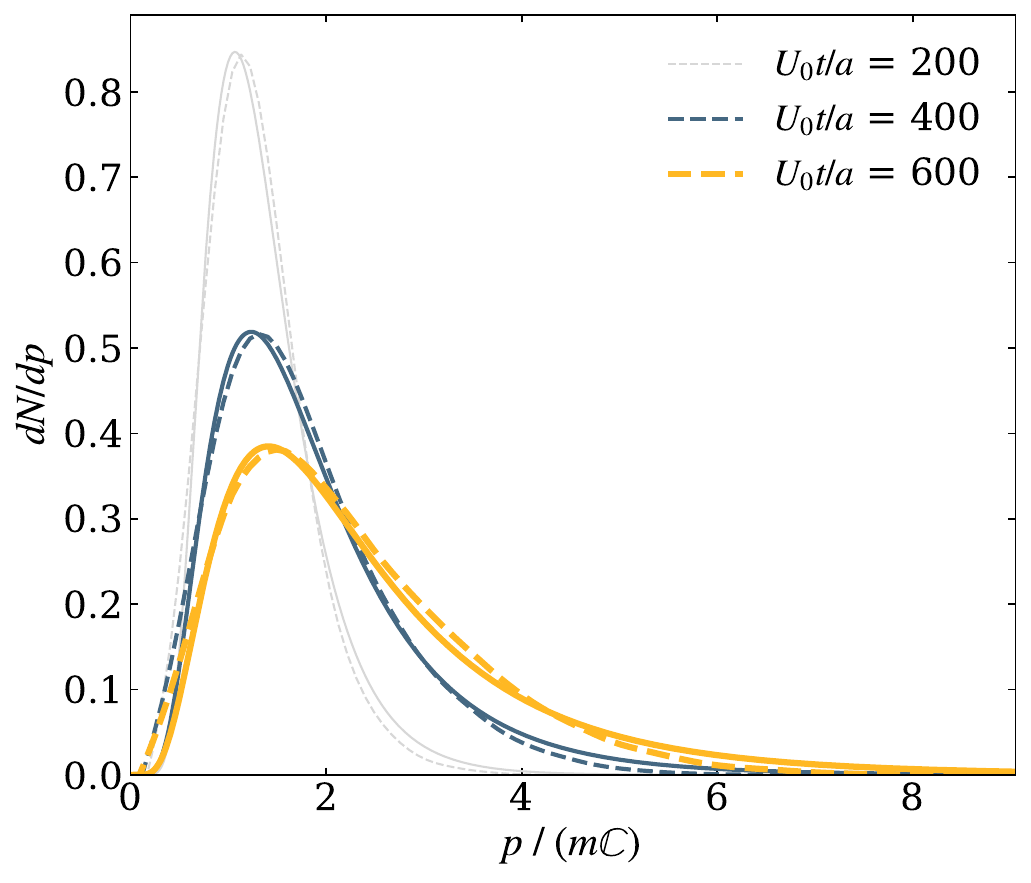}
    \caption{Particle momentum distribution of the population with $N_{\rm cross} > 40$, compared with log-normal fits from the GBM model (Equation \ref{log_normal}). The agreement supports the GBM description and its associated Fokker-Planck equation with $D_p \propto p^2$ (Appendix \ref{FP}).}
    \label{fig:GBM_pdf}
\end{figure}

Figure \ref{fig:GBM_pdf} demonstrates that GBM consistently captures the overall energy evolution throughout the acceleration process. Early in the simulation, the log-normal distribution matches the particle spectrum across all momenta. At later times, high-momentum particles maintain the log-normal distribution, while mild deviations emerge below the initial momentum $p_0$, where the relative diffusion rate $D_p/p^2$ of the measured diffusion coefficient increases toward small momenta (Appendix \ref{FP}). Despite this low-momentum excess, the GBM model successfully describes the stochastic particle acceleration in this system. The log-normal form also accounts for the hard effective index of Section \ref{spectrum}: the local slope of Equation \ref{log_normal} is $-d\ln f/d\ln p = 1 + \left[\ln(p/p_0) - \left(\xi - \sigma^2/2\right)t\right]/(\sigma^2 t)$, which lies below unity for momenta beneath the distribution's median and, at fixed $p > p_0$, decreases with time as the distribution broadens and advances. An apparent index harder than $\alpha = 1$ at intermediate momenta is therefore a generic transient feature of number-conserving diffusive acceleration without injection or escape, rather than a steady-state Fermi index.

\section{DISCUSSION} \label{discussion}

Understanding the transition to astrophysically relevant regimes remains crucial for applications to AGN jet boundaries, accretion disks, and ICM turbulence, where shear-driven turbulence naturally occurs. The detection of extended X-ray emission from kilo- to megaparsec-scale AGN jets \citep{harris_x-ray_2006} favors a synchrotron origin \citep{the_hess_collaboration_resolving_2020}. However, the short synchrotron cooling timescale requires a distributed \textit{in situ} (re-)acceleration mechanism to maintain the observed emission \citep{liu_particle_2017,wang_particle_2023}. Both Monte Carlo simulations \citep{kimura_ultrahigh-energy_2018} and relativistic MHD studies \citep{wang_particle_2023} have demonstrated stochastic shear acceleration as a promising mechanism for generating ultrahigh-energy cosmic rays consistent with observed X-ray emission.

Our fiducial simulations, which initialize particles with gyroradii $r_{\rm c,0} \gtrsim \Delta$ (where $\Delta = 2\,a$ is the shear layer width), effectively model the reacceleration phase in such jet boundary layers. Although the bulk flow of a jet is supersonic with respect to the ambient medium, the boundary shear develops against the cocoon of shocked gas that envelops the jet \citep{begelman_overpressured_1989}, which can be hot enough that its sound speed approaches the relativistic value $c/\sqrt{3}$; the velocity jump across the interface is then trans- or even subsonic with respect to the gas being stirred, closer to the regime simulated here than the bulk Mach number would suggest. For a jet with magnetic field strength $B \sim 10^{-2}~{\rm G}$ and shear layer width $\Delta \sim 0.1~{\rm pc}$ \citep{rieger_introduction_2019}, protons with Lorentz factors $\gamma \sim 10^9$ (corresponding to sub-EeV energies) have $r_{\rm c} \sim \Delta$, matching our simulation regime. Such particles can undergo continuous acceleration until their gyroradii exceed the jet width, at which point they escape the acceleration region (i.e., the Hillas limit). Although our simulations operate in the non-relativistic regime ($\mathbb{C} = 50\,U_0$), the fundamental acceleration mechanism, which relies on the geometry of orbit distortion rather than the flow speed itself, should persist in (trans-)relativistic flows. While our current setup does not capture relativistic fluid effects, the amplified motional electric fields $E \sim uB/c$ may enhance acceleration efficiency further.

The subsonic regime of our simulations is realized more directly in several non-relativistic environments. Cold clouds embedded in hot galactic winds and coronae provide one such site: clouds survive and grow when radiative cooling is efficient \citep{gronke_growth_2018}, and their interfaces host Kelvin-Helmholtz-driven turbulent mixing layers \citep{fielding_multiphase_2020} whose structure is set by the competition between turbulent mixing and the cooling of the mixed gas at $T \sim 10^5~{\rm K}$ \citep{fielding_multiphase_2020, tan_radiative_2021}. The sheets in which the entrained hot gas cools are thin, of order the cooling length $v\,t_{\rm cool}$, which for turbulent velocities $v \sim 10~{\rm km~s^{-1}}$ decreases from $\sim 0.2~{\rm pc}$ to $\sim 0.02~{\rm pc}$ as the mixed-gas density rises from $0.1$ to $1~{\rm cm^{-3}}$, while the velocity transition is spread over the multiphase region, whose width is of order a parsec for clouds of $\sim 10~{\rm pc}$ \citep{tan_radiative_2021}. For $B \sim 1~\mu{\rm G}$, protons in the regime simulated here ($r_{\rm c} \sim \Delta$) have $E \sim 10^{13}$--$10^{14}~{\rm eV}$ when $\Delta$ is identified with the cooling length and $E \sim 10^{15}~{\rm eV}$ when $\Delta$ is identified with the width of the multiphase region, so the ambient Galactic cosmic ray population supplies the pre-existing energetic population that our setup assumes (Section \ref{setup}; \citealt{blasi_origin_2013}), and no injection mechanism is required. The drift fitted in Section \ref{acceleration} corresponds to an acceleration time $t_{\rm acc} = 1/\xi \approx 7\,(\mathbb{C}/U_0)^2\,\Delta/\mathbb{C}$ in the simulation; because the energy gain per crossing scales as $(U_0/\mathbb{C})^2$ (Equation \ref{mean_change}; Figure \ref{fig:second_order}) and the crossing rate as $\mathbb{C}/\Delta$, the physical value follows by replacing $\mathbb{C}$ with $c$, $t_{\rm acc} \approx 8~{\rm Myr}\,(\Delta/10^{-2}~{\rm pc})\,(2U_0/100~{\rm km~s^{-1}})^{-2}$, where $2U_0$ is the velocity difference across the layer, so the energization is fastest for thin layers in fast winds. A quantitative assessment, however, requires the cooling that sets both scales and the multiphase structure, neither of which our isothermal setup includes.

Galaxy clusters host subsonic shear layers on far larger scales and provide evidence that such layers persist, although their widths place them outside the regime simulated here. Cold fronts carry tangential shear at or below the sound speed of the surrounding medium \citep{markevitch_shocks_2007}, with sloshing maintaining the shear over long timescales \citep{roediger_kelvinhelmholtz_2013}: XRISM measurements of the prototypical front in Abell 3667 resolve a line-of-sight velocity change of $\sim 500~{\rm km~s^{-1}}$ across the interface \citep{omiya_xrism_2026}, and candidate Kelvin-Helmholtz rolls have been identified along sloshing fronts \citep{walker_is_2017}, although sufficiently strong draped magnetic fields can suppress the instability locally \citep{omiya_xrism_2026}, consistent with the stabilization threshold discussed in Section \ref{turbulence}. The tails of cluster radio galaxies, bent by the motion of their hosts through the intracluster medium at speeds comparable to the sound speed, show gentle \textit{in situ} re-energization \citep{de_gasperin_gentle_2017}, but of radio-emitting electrons with $\gamma \sim 10^3$--$10^4$, whose gyroradii in microgauss fields are $\sim 10^{-6}~{\rm pc}$, many orders of magnitude below any plausible width of the tail boundary. The regime simulated here would instead apply to protons with $r_{\rm c} \sim \Delta$, that is, $E \sim 2 \times 10^{17}~{\rm eV}\,(B/2~\mu{\rm G})\,(\Delta/100~{\rm pc})$, and is viable only if the boundary layer is well below a kiloparsec wide and the tails carry such a population. More generally, cold fronts are kiloparsec-scale interfaces \citep{markevitch_shocks_2007}, so all but the most energetic cluster cosmic rays have $r_{\rm c} \ll \Delta$: protons of $1$--$10^3~{\rm GeV}$ in microgauss fields have $r_{\rm c} \sim 10^{-6}$--$10^{-3}~{\rm pc}$. The acceleration mechanism nevertheless persists below the fiducial ordering. The runs in Appendix \ref{extrapolation} extend down to $r_{\rm c,0} \simeq 0.2\,\Delta$ and show sustained energization, so the mechanism is not fine-tuned to $r_{\rm c} \sim \Delta$; the gradual-shear limit proper \citep{rieger_shear_2004}, in which the particle mean free path is much smaller than the layer width and which applies to these systems, remains untested here.

The kinetic signatures of this acceleration mechanism closely resemble those found in different physical contexts, such as PIC studies of magnetically dominated relativistic turbulence. For instance, \citet{comisso_interplay_2019} found that high-energy particles in that regime develop pronounced pitch angle anisotropy, with strong probability peaks near $\mu = \pm 1$ at moderate energies and progressive enhancement near $\mu = 0$ at higher energies. They attributed this anisotropy to the transition between acceleration mechanisms: reconnection-mediated parallel electric fields $E_\parallel$ align injected particles with the magnetic field, while motional electric fields $E_\perp$ propel high-energy particles perpendicular to it. Our simulations in non-relativistic shear-driven turbulence, where reconnection plays no role in particle energization, nevertheless reproduce qualitatively similar pitch angle evolution (Figure \ref{fig:pitch_dist}). The emergence of similar pitch angle signatures across these disparate regimes suggests that pitch angle anisotropy may be a generic feature of turbulent acceleration whenever perpendicular electric fields dominate particle energization.

Furthermore, the microscopic mechanism of orbit distortion identified here (Figure \ref{fig:distortion}) provides a direct physical validation of the effective theory for stochastic acceleration recently proposed by \citet{2025PhRvE.112a5205L}. In this generalized framework, the advection coefficient in momentum space is shown to be strictly positive ($A_p > 0$) due to a noise-induced drift of the stochastic process, where the randomly oriented electric fields $\delta \boldsymbol{E}$ preferentially focus the spatial component of the particle momentum toward directions that maximize energy gain (i.e., toward $q \delta \boldsymbol{E}$). This implies that the energy gain from favorable interactions exceeds the loss from unfavorable ones, resulting in a net energy drift of order $\mathcal{O}(v_E^2/c^2)$, where $\boldsymbol{v}_E$ denotes the velocity of the local comoving frame in which the electric field vanishes. This theoretical prediction is kinematically realized in our simulations via the path length asymmetry shown in Figure \ref{fig:distortion}: the electric force consistently extends the interaction path during accelerating phases ($\boldsymbol{v} \cdot \boldsymbol{E} > 0$) and shortens it during decelerating phases ($\boldsymbol{v} \cdot \boldsymbol{E} < 0$). The kinematic asymmetry introduces a net bias to the stochastic interactions with the electric fields, giving rise to the second-order scaling with shear velocity ($\langle \Delta \epsilon \rangle \propto U_{\rm shear}^2$) demonstrated in Figure \ref{fig:second_order}. Consequently, our results confirm that second-order Fermi acceleration in shear flows proceeds through the systematic rectification of stochastic interactions with motional electric fields generated by the turbulence.

In this work, we assume continuously driven turbulence with sustained shear profiles, which is directly relevant to astrophysical environments where external forcing maintains velocity gradients over extended timescales, such as AGN jets where the central engine continuously drives the flow, or accretion disk boundary layers where the differential rotation persists. However, in environments dominated by supersonic turbulence, such as molecular clouds or the warm ISM, shear layers may be transient substructures that form and decay on eddy turnover times. Recent work by \citet{liangHybridSimulationsSupersonic2025, liangHybridSimulationsSupersonicViscosity2026} has explored this complementary regime of freely decaying transonic/supersonic shear using hybrid PIC simulations, addressing particle acceleration in the first paper and the backreaction of cosmic rays on the shear in the second. They also find minimal particle acceleration in decaying subsonic turbulence, consistent with Figure \ref{fig:steady_state}. Nevertheless, in the supersonic regime, non-thermal particle acceleration can be achieved without requiring a sustained driving force. The shocklets are able to accelerate particles efficiently and produce significant non-thermal fractions before the free energy depletes, with larger sonic Mach number $M_s$ leading to larger acceleration out of the thermal pool \citep{liangHybridSimulationsSupersonic2025}. Pre-existing energetic particles, in turn, facilitate the dissipation of the shear by carrying momentum between layers, an effective cosmic ray viscosity that grows with the particle energy density provided the gyroradii remain smaller than the shear scale \citep{liangHybridSimulationsSupersonicViscosity2026}. Together, these works span the realistic astrophysical parameter space, with persistent large-scale flows favoring sustained acceleration while turbulent substructures enable transient acceleration.

A further idealization is the isothermal equation of state, which removes the heat that turbulent dissipation would otherwise deposit in the shear layers. In our subsonic, high-$\beta$ regime, retaining this heat would mainly raise the sound speed, further reducing the compressibility of the flow while leaving the vortex stretching and magnetic field line folding that generate the motional electric fields essentially unchanged. Quantifying thermal feedback, including regimes where heating or radiative cooling shapes the layer structure, requires evolving an explicit energy equation and remains beyond the scope of this study.

Our simulations are two-dimensional, which restricts particle transport and the structure of the turbulence. With one ignorable coordinate, particles remain tied to the magnetic field lines they start on \citep{jones_charged-particle_1998}, so excursions across the mean field are set by the gyroradius rather than by cross-field transport. Cross-field transport is decisive where particles must return against an advecting flow, as in the injection of ions at quasi-perpendicular shocks \citep{orusa_fast_2023}. No such competition exists in our simulations, because the layers are held in place by the stirring force and particles with $r_{\rm c,0} \gtrsim \Delta$ near them cross them from the outset. In three dimensions, field-line wandering and drifts would let guiding centers migrate into and out of the layers, redistributing which particles are accelerated and for how long, without changing how they gain energy at each encounter.

The turbulence itself would also differ. In three dimensions the electric field would lose coherence along $z$ over a finite distance, so the coherent segments of Section~\ref{acceleration} would be bounded by the shorter of the layer transit time and the time to cross that distance. Three-dimensional simulations of a driven, weakly magnetized double shear layer closely analogous to ours indicate that this distance is long compared with the layer width. The turbulence remains strongly anisotropic, the Kelvin-Helmholtz eddies first form as cylinders with almost no variation along the coordinate that is ignorable in our setup, and the energetically dominant three-dimensional structures that develop later are jets invariant along the flow and varying along that coordinate on the scale of the domain, with no appreciable buildup of small-scale structure along it \citep{tripathi_large-scale_2026}, so the transit time would remain the limiting timescale. Shorter segments would in any case lower the energy change per segment and hence the acceleration rate, but not the asymmetry that produces the net gain: the mean and mean-square energy changes per segment (Equations \ref{W0_square} and \ref{W1_mean}) scale alike with the segment duration, so their ratio, which sets the second-order character of the process, is unchanged, as is the quadratic dependence on $U_{\rm shear}$, which enters only through $E_z$. A three-dimensional cascade would also carry more power to small scales. In the driven, statistically stationary layers considered here, the large-scale energy reservoirs that the particles sample are maintained by the stirring force against dissipation, so additional power at small scales should not by itself alter the energization. Preliminary thin-slab experiments ($L_z = L_x/10$) indicate that particle energization remains dominated by out-of-plane electric fields, with only modest contributions from in-plane components, suggesting that the fundamental acceleration mechanism identified here should persist in fully three-dimensional configurations.

Finally, we address the dynamical impact of the accelerated particles on the background flow. In our simulations, we observe modest particle backreaction: the power absorbed by particles constitutes approximately $10\%$ of the power injected by the external stirring force, and, relative to an otherwise identical simulation evolved without backreaction (i.e., in the test-particle regime), both the turbulent energy spectrum and the volume-averaged kinetic and magnetic energy densities remain statistically unchanged (Section~\ref{turbulence}). Because the backreaction force is proportional to the particle mass density, lighter particle loadings would couple to the flow even more weakly. This modest feedback contrasts with recent studies of cosmic-ray-modified turbulence, where diffusive cosmic rays significantly damp compressive modes and truncate the turbulent cascade \citep{bustard_cosmic-ray_2023, habegger_cosmic-ray_2024}. This discrepancy arises from the distinct nature of the turbulence and the particle transport regime in our setup. While those works focus on compressive turbulence damped by low-energy, highly diffusive cosmic rays, our shear-driven turbulence is dominated by solenoidal vortex motions. Furthermore, the particles in our simulations have larger gyroradii ($r_{\rm c} \gtrsim \Delta$) and lower scattering rates compared to the fluid regime modeled by \citet{bustard_cosmic-ray_2023} and \citet{habegger_cosmic-ray_2024}. As a result, our particles interact stochastically with the shear layers, preventing the tight hydrodynamical coupling required for strong viscous damping.

\section{Conclusions} \label{conclusions}
In this work, we present the first numerical study of particle acceleration in sustained, subsonic, magnetized, non-relativistic turbulence driven solely by velocity shear, including full particle backreaction. Using two-dimensional MHD-PIC simulations in the ideal MHD limit, we demonstrate that sustained particle acceleration occurs exclusively through the ideal motional electric field. Our configuration with an initially flow-aligned magnetic field enables particle acceleration via the KHI alone, distinct from the combined effects of KHI and DKI reported by \citet{tsung_dissipation_2025}. In our fiducial simulations, we bypass the need for reconnection-mediated injection by initializing particles with sufficient energy to cross turbulent layers from the outset. Nevertheless, particles can still undergo sustained acceleration when their initial gyroradii are a fraction of the shear layer width, down to $r_{\rm c,0} \simeq 0.2\,\Delta$ (see Appendix \ref{extrapolation}). By employing a stirring force to maintain the shear profile against dissipation, we achieve quasi-stationary turbulence and obtain the following findings:

\begin{enumerate}
    \item \textbf{Sustained driving is required for continuous particle acceleration in subsonic shear-driven turbulence.} In addition to observing the expected decay of turbulence without energy injection, we show that the stirring force maintains stable energy reservoirs that enable linear growth of average particle energy throughout the simulation, whereas freely decaying turbulence results in saturation. These reservoirs are statistically unchanged when particle backreaction is switched off, so this conclusion does not rest on the substantial particle energy density adopted here.
    
    \item \textbf{Initially monoenergetic particles develop significant non-thermal populations.} We demonstrate that non-thermal tails emerge in this non-relativistic and subsonic regime. The particle spectrum broadens significantly, with the effective slope of the non-thermal component flattening in time as acceleration proceeds.
    
    \item \textbf{Orbit distortion is identified as the driver of net energy gain.} We provide direct microscopic evidence that electric fields distort particle gyro-orbits, leading to an asymmetry in the interaction path length. Acceleration phases extend the interaction path along the electric force, increasing the energy gain, while deceleration phases shorten the path, reducing the energy loss. Our results confirm that this kinematic asymmetry leads to systematic energization.

    \item \textbf{Mean energy change scales quadratically with shear velocity.} We numerically validate that the measured scaling $\langle\Delta\epsilon\rangle \propto U_{\text{shear}}^{2.01\pm0.02}$ matches the theoretical expectation for second-order Fermi acceleration, arising from the rectification of stochastic interactions with turbulent electric fields.
    
    \item \textbf{Particle acceleration follows geometric Brownian motion.} We show that the process is approximately described by geometric Brownian motion, in which the systematic drift is the accumulated second-order bias of the stochastic encounters and is comparable in strength to the diffusion ($\xi/\sigma^2 \approx 2.4$). This produces log-normal distributions for particles frequently crossing shear layers and confirms that the process is a momentum diffusion whose drift is generated by the fluctuations themselves.
\end{enumerate}

These findings demonstrate that magnetized shear-driven turbulence excited by KHI alone can efficiently accelerate particles through second-order Fermi processes. Future work should explore the effective viscosity induced by accelerated particles and how this backreaction modifies the background flow as a function of particle energy density (see \citealt{liangHybridSimulationsSupersonicViscosity2026} for the decaying supersonic case). Three-dimensional simulations would capture additional physics including the twisting and folding of out-of-plane magnetic field that may generate strong in-plane electric fields, and would test the expectation of Section~\ref{discussion} that the restrictions specific to two dimensions, the tying of particles to their field lines and the coherence of the electric field along $z$, alter the population statistics and the acceleration rate rather than the mechanism itself. Such extensions would further clarify the bidirectional coupling between shear-driven turbulence and cosmic rays: turbulence accelerates particles while the accelerated particles, in turn, modify the flow through momentum transfer across shear layers.

\begin{acknowledgments}
This work used the Delta system at the National Center for Supercomputing Applications [award OAC 2005572] through allocation PHY250299 from the Advanced Cyberinfrastructure Coordination Ecosystem: Services \& Support (ACCESS) program, which is supported by National Science Foundation grants \#2138259, \#2138286, \#2138307, \#2137603, and \#2138296. The authors acknowledge the Texas Advanced Computing Center (TACC) at The University of Texas at Austin for providing computational resources that have contributed to the research results reported within this paper. This work also used the Stampede3 system at TACC [award \#2320757] through ACCESS allocation PHY250146. MR acknowledges support from the National Science Foundation Collaborative Research Grant NSF AST-2009227 and NASA ATP grant 80NSSC23K0014. EZ acknowledges support by NSF PHY-2409224 and the Simons Collaboration on Extreme Electrodynamics of Compact Sources. DC was partially supported by NASA grant 80NSSC18K1726, NSF grants AST-2510951 and AST-2308021. XS acknowledges support by Multimessenger Plasma Physics Center (MPPC, NSF grant PHY-2206607). This work was performed in part at the Kavli Institute for Theoretical Physics (KITP) during the ``Turbulence in Astrophysical Environments" program, supported in part by the NSF PHY-2309135 grant to the KITP, and at the Aspen Center for Physics (ACP) during the ``Cosmic Ray Feedback in Galaxies and Galaxy Clusters" summer program. The ACP is supported by the National Science Foundation grant PHY-2210452, and by grants from the Simons Foundation (1161654, Troyer) and Alfred P. Sloan Foundation (G-2024-22395). Finally, we thank the reviewer for a careful and constructive report, which has improved both the analysis and presentation of this paper.
\end{acknowledgments}

\begin{appendix}

\section{Numerical Details}\label{details}

The background flow adopts an isothermal equation of state with sound speed $c_s = 5\,U_0$ (Section \ref{setup}).

To seed the Kelvin-Helmholtz instability, we applied small-amplitude random perturbations to the initial velocity field. The perturbed velocity components are given by:
\begin{align}
    u_x(x,y) &= u_{x,0}(y) + \delta u \cdot \mathcal{R}_x(x,y), \\
    u_y(x,y) &= \delta u \cdot \mathcal{R}_y(x,y), \\
    u_z(x,y) &= 0,
\end{align}
where $u_{x,0}(y)$ is the background shear profile defined in Equation \ref{shear_velocity}, $\delta u$ is the perturbation amplitude, and $\mathcal{R}_{x,y}(x,y)$ are spatially varying random numbers uniformly distributed in $[-0.5, 0.5]$. The perturbation amplitude is set to $\delta u = 0.001\, U_0$ to keep the initial state in the linear regime while allowing sufficient perturbation growth.

The initial particle velocities are first assigned isotropically in the local fluid rest frame: 
\begin{equation}
    v_x' = v_0' \cos\theta, \quad v_y' = v_0' \sin\theta \cos\phi, \quad v_z' = v_0' \sin\theta \sin\phi,
\end{equation}
where $\theta = \arccos(2\mathcal{R}_1 - 1)$ and $\phi = 2\pi \mathcal{R}_2$. The initial velocity magnitude in the fluid frame is set to $v_0' = \left( p'/m \right)_0 / \gamma_0'$, with $\left( p'/m \right)_0 = \mathbb{C}$ corresponding to $\gamma_0' = \sqrt{2}$. Averaging the individual gyroradii $\left( p_{\perp}/m \right) / \left[ (q/mc)\, B_0 \right]$ over the isotropic distribution yields $r_{\rm c,0} = (\pi/4) \left( p'/m \right)_0 / \left[ (q/mc)\, B_0 \right]$, where the factor $\pi/4$ is the average of $\sin\theta$ over the isotropic distribution, and the charge-to-mass ratio is chosen such that $r_{\rm c,0} \approx 2\,a$ matches the shear layer width. Because the Lorentz boost modifies only the momentum component along $\boldsymbol{B}_0$, $r_{\rm c,0}$ is independent of the shear profile. $\mathcal{R}_1, \mathcal{R}_2$ are independent random numbers uniformly distributed in $[0, 1]$. 

To obtain velocities in the laboratory frame where the fluid has velocity $\boldsymbol{u}_{\rm shear} = u_{x,0}(y)\,\hat{\boldsymbol{x}}$, we apply a Lorentz boost to particles. The transformation of particle spatial four-velocity yields:
\begin{equation}
    \boldsymbol{u} = \boldsymbol{u}' + \left[ (\gamma_{\rm shear} - 1)\frac{\boldsymbol{u}' \cdot \boldsymbol{u}_{\rm shear}}{u_{\rm shear}^2} + \gamma_0' \right]\boldsymbol{u}_{\rm shear},
\end{equation}
where $\boldsymbol{u} = \gamma \boldsymbol{v} = \boldsymbol{p}/m$ represents particle spatial four-velocity in the lab frame. $\gamma_{\rm shear} = (1 - u_{\rm shear}^2/\mathbb{C}^2)^{-1/2}$ is the Lorentz factor of the shear flow. This prescription ensures that particles have an isotropic velocity distribution in the local fluid rest frame while properly accounting for the shear profile.

\begin{figure}
    \centering
    \includegraphics[width=1.0\linewidth]{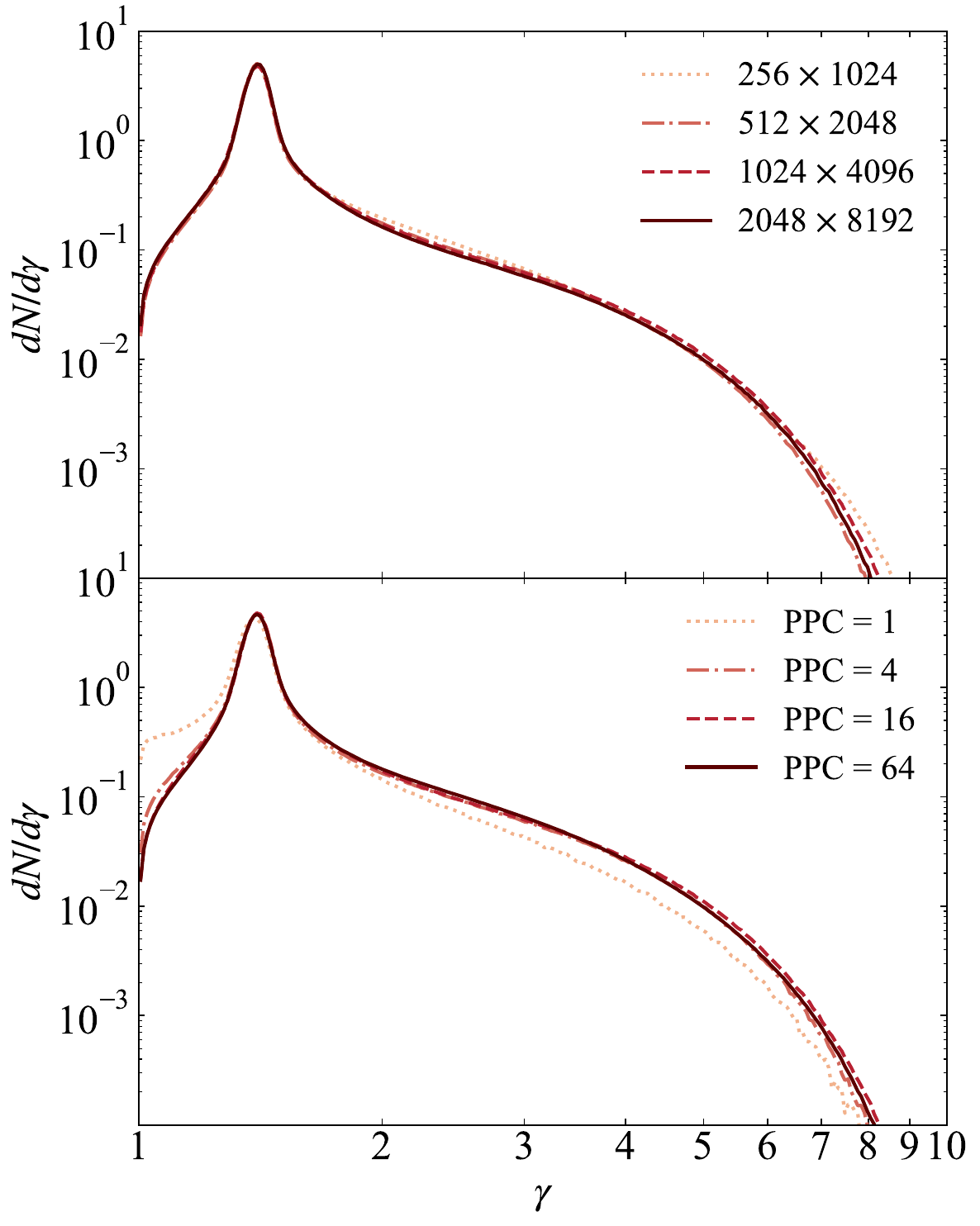}
    \caption{Convergence of particle energy spectra with simulation parameters. \textbf{Top:} Particle energy distribution, with curves color-coded by the number of grid cells while fixing the total number of particles. \textbf{Bottom:} Particle energy distribution, with curves color-coded by particles-per-cell (PPC) while fixing the spatial resolution.} 
    \label{fig:convergence}
\end{figure}

We conducted convergence tests to verify the robustness of our results. Figure \ref{fig:convergence} shows particle energy spectra measured at $t = 600\,a/U_0$ for different spatial resolutions and particle statistics. The top panel demonstrates convergence with respect to grid resolution, comparing simulations with resolutions from $256 \times 1024$ to $2048 \times 8192$ cells while maintaining a fixed total particle number. These resolutions correspond to initial particle gyroradii ranging from $r_{\rm c,0} \approx 16\Delta x$ to $130\Delta x$, where $\Delta x$ denotes the grid cell size. The energy distributions exhibit good agreement across all resolutions in the thermal peak, non-thermal tail, and high-energy cutoff, with only minor deviations appearing in the lowest-resolution run at $\gamma \gtrsim 6$.

The bottom panel tests convergence with respect to particle statistics by varying the particles-per-cell (PPC) from 1 to 64 while fixing the spatial resolution at $1024 \times 4096$. The distributions converge well for PPC $\geq 4$, indicating that adequate particle statistics are achieved beyond this threshold. The PPC = 1 case shows significant noise at high energies due to poor sampling statistics, while PPC $\geq 4$ cases produce consistent results. Based on these tests, we adopt a fiducial resolution of $1024 \times 4096$ cells with PPC = 16 for the simulations presented in this work, balancing computational cost with numerical accuracy. We have additionally verified convergence with respect to temporal resolution by comparing runs with different timestep sizes (not shown).

We also verified that numerical heating is negligible in our setup. In PIC simulations, numerical heating of the particles can appear due to finite-particle shot noise generating spurious electromagnetic fluctuations. The impact of this artifact typically scales inversely with PPC. However, the particle energy spectra converge for PPC $\ge 4$ (Figure \ref{fig:convergence}, bottom panel) and do not exhibit the artificial spectral hardening at low PPC characteristic of numerical heating. Furthermore, the saturation of particle energy density in the freely decaying turbulence (Figure \ref{fig:steady_state}, bottom panel) confirms the robustness of particle energization; if the heating were numerical, energy would continue to rise linearly even as physical turbulence decays.

\section{Particle-Field Interaction}\label{derivation}
To explain the microscopic acceleration mechanism discussed in Section \ref{acceleration}, we consider the motion of a relativistic particle with charge $q$, mass $m$, and Lorentz factor $\gamma$ interacting with a local electric field $\boldsymbol{E}$ and background magnetic field $\boldsymbol{B}$. The equation of motion is:
\begin{equation}
\frac{d\boldsymbol{p}}{dt} = q(\boldsymbol{E} + \boldsymbol{v}\times\boldsymbol{B}/c),
\end{equation}
where $\boldsymbol{p} = \gamma m \boldsymbol{v}$ is the particle momentum. We decompose the particle velocity into an unperturbed component $\boldsymbol{v}_0$ (governed solely by the magnetic field) and a perturbation $\delta\boldsymbol{v}$ induced by the electric field: $\boldsymbol{v}(t) = \boldsymbol{v}_0(t) + \delta\boldsymbol{v}(t)$.

Assuming the interaction time $\tau$ is much shorter than the gyroperiod, we can neglect the magnetic force on the perturbation. The momentum perturbation integrated over $\tau$ is approximately $\delta\boldsymbol{p} \approx q\boldsymbol{E}\tau$. The corresponding velocity perturbation is derived from the relativistic identity $d\boldsymbol{v} = (\mathbb{C}^2/\epsilon)[d\boldsymbol{p} - \boldsymbol{\beta}(\boldsymbol{\beta}\cdot d\boldsymbol{p})]$, where $\epsilon = \gamma m \mathbb{C}^2$ is the particle energy and $\boldsymbol{\beta} = \boldsymbol{v}/\mathbb{C}$:
\begin{equation}
\delta\boldsymbol{v} \approx \frac{q\tau}{\gamma m} \left[ \boldsymbol{E} - \boldsymbol{\beta}_0 (\boldsymbol{\beta}_0 \cdot \boldsymbol{E}) \right].
\end{equation}
This velocity perturbation leads to an effective spatial displacement $\delta\boldsymbol{x}$ in the direction of the electric force. Integrating $\delta \boldsymbol{v}$ over time $\tau$:
\begin{equation}
\delta\boldsymbol{x} = \int_0^\tau \delta\boldsymbol{v} \, dt \approx \frac{q\tau^2}{2\gamma m} \left[ \boldsymbol{E} - \boldsymbol{\beta}_0 (\boldsymbol{\beta}_0 \cdot \boldsymbol{E}) \right],
\end{equation}
which introduces an asymmetry in the interaction path length. In aligned segments ($\boldsymbol{v}_0 \cdot \boldsymbol{E} > 0$), the particle accelerates, and $\delta\boldsymbol{x}$ extends the path length over which the electric force acts. Conversely, in anti-aligned segments ($\boldsymbol{v}_0 \cdot \boldsymbol{E} < 0$), the particle decelerates, shortening the path length. This asymmetry, confirmed by Figure \ref{fig:distortion}, makes the work done during acceleration phases exceed the energy lost during deceleration phases.

We calculate the total work done over the interval $\tau$ as $\Delta\epsilon = \int_0^\tau q \boldsymbol{E} \cdot (\boldsymbol{v}_0 + \delta\boldsymbol{v}) \, dt$. This splits into a linear term and a second-order distortion term:
\begin{equation}
\Delta\epsilon \approx \underbrace{\int_0^\tau q \boldsymbol{E} \cdot \boldsymbol{v}_0 \, dt}_{W_0} + \underbrace{\int_0^\tau q \boldsymbol{E} \cdot \delta\boldsymbol{v} \, dt}_{W_1}.
\end{equation}
The linear term $W_0$ averages to zero over many stochastic interactions ($\left \langle W_0 \right \rangle = 0$) but acts as the primary driver of energy diffusion. The systematic energy gain arises entirely from the second-order term $W_1$. Substituting $\delta \boldsymbol{v}$ into the work integral and assuming $\boldsymbol{E}$ is approximately constant over $\tau$, we find:
\begin{equation}
    W_1 \approx \frac{q^2 E^2 \tau^2}{2\gamma m} \left( 1 - \beta_0^2 \cos^2\theta \right),
\end{equation}
where $\theta$ is the angle between the unperturbed velocity $\boldsymbol{v}_0$ and the electric field $\boldsymbol{E}$. Since $\beta_0 < 1$ and $\cos^2\theta \le 1$, the term $(1 - \beta_0^2 \cos^2\theta)$ is strictly positive, ensuring $W_1 > 0$.

We can quantify the statistical properties of the energy change $\Delta\epsilon = W_0 + W_1$. The squared energy change is dominated by the first-order term:
\begin{equation}
    (\Delta \epsilon)^2 = W_0^2 + 2W_0W_1 + W_1^2 \approx W_0^2,
\end{equation}
where we have neglected the higher-order terms $2W_0W_1$ (which eventually averages to zero due to the random sign of $W_0$) and $W_1^2$ (which is fourth-order in $E$). Thus, the mean squared energy change is given by:
\begin{equation}\label{W0_square}
    \left \langle (\Delta \epsilon)^2 \right \rangle \approx \left \langle W_0^2 \right \rangle \approx \left \langle q^2 v_0^2 E^2 \tau^2 \cos^2\theta \right \rangle.
\end{equation}
In contrast, the mean energy change is determined solely by the second-order distortion term $W_1$, as $\left \langle W_0 \right \rangle = 0$:
\begin{equation}\label{W1_mean}
    \left \langle \Delta \epsilon \right \rangle = \left \langle W_1 \right \rangle = \left\langle \frac{q^2 E^2 \tau^2}{2 \gamma m} \left( 1 - \beta_0^2\cos^2\theta \right) \right\rangle.
\end{equation}
Since $\left \langle \Delta\epsilon \right \rangle$ is always positive, this result demonstrates that while the random alignment of $\boldsymbol{v}$ and $\boldsymbol{E}$ drives diffusion, the electric field-induced distortion of the particle orbit introduces a systematic bias, leading to net second-order Fermi acceleration.

These general expressions reduce to Equations (\ref{mean_square}) and (\ref{mean_change}) of the main text when specialized to the shear-crossing geometry of Section \ref{acceleration}. There, the electric field is dominated by its out-of-plane component ($\boldsymbol{E} \approx E_z\,\hat{\boldsymbol{z}}$; Figure \ref{fig:power}), so that $v_0 \cos\theta = v_z$ and $\beta_0 \cos\theta = \beta_z$, and the correlation time is the transit time of the layer half-width at the transverse velocity, $\tau \approx a/|v_y|$. Substituting these into Equations (\ref{W0_square}) and (\ref{W1_mean}) yields
\begin{equation}
\left\langle (\Delta \epsilon)^2 \right\rangle \approx \left\langle q^2 E_z^2 a^2\, \frac{v_z^2}{v_y^2} \right\rangle = \left\langle \epsilon_*^2\, \frac{v_z^2}{v_y^2} \right\rangle,
\end{equation}
\begin{equation}
\left\langle \Delta \epsilon \right\rangle \approx \left\langle \frac{\epsilon_*^2}{2 \gamma m\, v_y^2} \left( 1 - \beta_z^2 \right) \right\rangle = \left\langle \frac{\epsilon_*^2}{2 \epsilon}\, \frac{1 - \beta_z^2}{\beta_y^2} \right\rangle,
\end{equation}
where $\epsilon_* \equiv qE_z a$ as defined in Section \ref{acceleration}. For gyration about the flow-aligned field, $v_y$ and $v_z$ are the two components of the perpendicular velocity, $v_y = v_\perp \cos\psi$ and $v_z = -v_\perp \sin\psi$ with gyrophase $\psi$, so the ratios $v_z^2/v_y^2$ and $(1-\beta_z^2)/\beta_y^2$ are of order unity at typical phases. Segments with $|v_y| \to 0$ do not complete a crossing within a gyration; their duration is instead limited by the gyroperiod, which regularizes the gyrophase average of $v_z^2/v_y^2$. Note also that $\Omega \tau \sim (a/r_{\rm c})(v_\perp/|v_y|)$, so the neglect of velocity rotation during a segment holds at typical gyrophases for $r_{\rm c} \gtrsim a$, the regime of our fiducial setup. Absorbing the resulting order-unity factors into the scalings recovers Equations (\ref{mean_square}) and (\ref{mean_change}) of the main text.

\section{Fokker-Planck Equation}\label{FP}

The geometric Brownian motion (Equation \ref{SDE}) in It\^o form relates to the Fokker-Planck formalism through the Stratonovich convention, in which the drift $\mu_S$ and the It\^o drift differ by the correction $\frac{1}{2}\sigma\, d\sigma/dp$:
\begin{equation}
    dp_t = \left[ \mu_S + \frac{1}{2} \sigma \frac{d\sigma}{dp} \right] dt + \sigma\, dW_t,
\end{equation}
where $\mu_S$ is the drift beyond this correction. In our ideal MHD setup there are no parallel electric fields, so $\mu_S$ contains no deterministic acceleration. It is the systematic gain that accumulates from the second-order bias of individual stochastic encounters (Equations \ref{mean_square} and \ref{mean_change}; Appendix \ref{derivation}).

This corresponds to the Fokker-Planck equation:
\begin{align}\label{eq:FP}
\frac{\partial f}{\partial t} &= -\frac{\partial}{\partial p} \left[A_p f\right] + \frac{\partial^2}{\partial p^2} \left[D_p f\right], \\
A_p &= \mu_S + \frac{1}{2} \sigma \frac{d\sigma}{dp}, \\
\quad D_p &= \frac{\sigma^2}{2},
\end{align}
where $A_p$ and $D_p$ are the advection and diffusion coefficients. For $\sigma \propto p$, we obtain $A_p \propto p$ and $D_p \propto p^2$, yielding the log-normal distribution (Equation \ref{log_normal}); writing the noise amplitude as $\sigma p$ with the constant $\sigma$ of Equation \ref{SDE}, $A_p = \xi p$ and $D_p = \sigma^2 p^2/2$. The fitted coefficients of Figure \ref{fig:GBM_evo} give $\xi/\sigma^2 = A_p p/(2D_p) \approx 2.4$, so the drift is dominated by the accumulated bias rather than by the Stratonovich correction alone, which would give $\xi = \sigma^2/2 \approx 3\times10^{-4}\,U_0/a$, five times below the fitted value. A ratio $\xi/\sigma^2$ of order unity is what one expects when the drift is generated by the fluctuations themselves: Equations \ref{mean_square} and \ref{mean_change} give $\langle\Delta\epsilon\rangle \approx \langle(\Delta\epsilon)^2\rangle/\epsilon$ per encounter, that is, an advection coefficient of order $D_p/p$, or $\xi$ of order $\sigma^2$.

\begin{figure}
    \centering
    \includegraphics[width=1.0\linewidth]{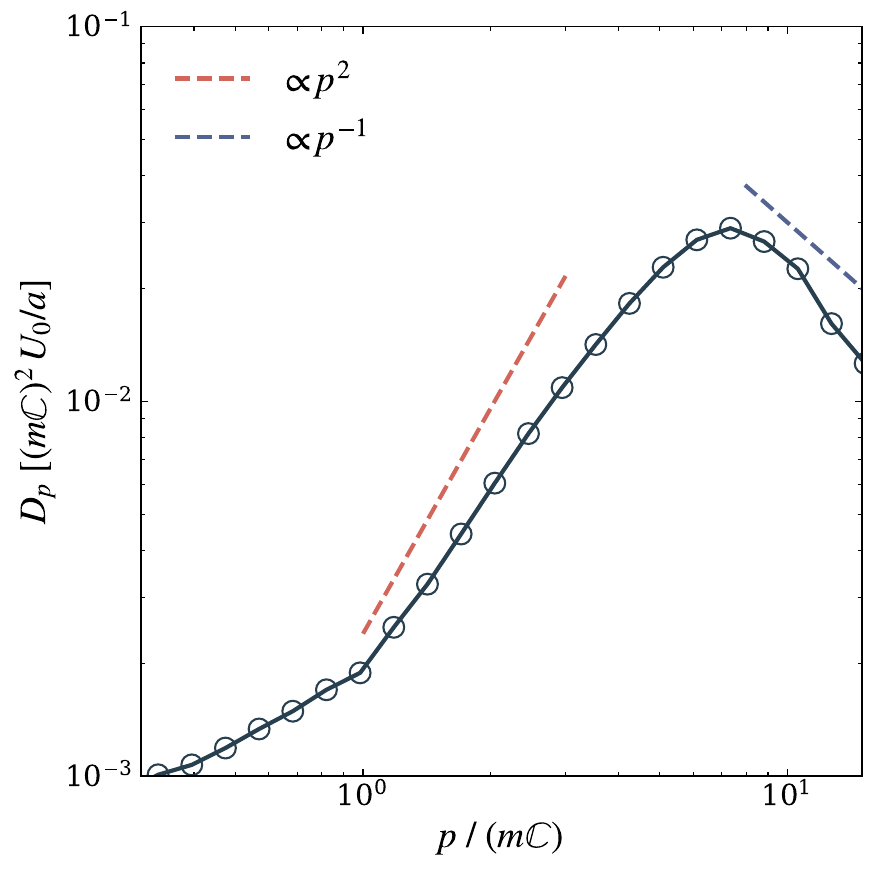}
    \caption{Momentum diffusion coefficient $D_p$ versus particle momentum. Circles: simulation data using $D_p = \langle(\Delta p)^2\rangle/(2\Delta t)$. Dashed lines: $\propto p^2$ (red; the ultrarelativistic GBM scaling) and $\propto p^{-1}$ (blue) scalings.}
    \label{fig:diffusion}
\end{figure}

We measured $D_p = \langle (\Delta p)^2 \rangle/(2\Delta t)$ from simulation data (Figure \ref{fig:diffusion}). At low momenta, the diffusion coefficient $D_p$ increases with particle momentum, with a local slope shallower than the GBM value of 2, so the relative diffusion rate $D_p/p^2$ grows toward small momenta, consistent with the mild low-momentum excess relative to the log-normal form in Figure \ref{fig:GBM_pdf}. In the trans-relativistic range ($1 \lesssim p/(m\mathbb{C}) \lesssim 7$), the steep rise is consistent with an energy diffusion coefficient $D_\epsilon \equiv \langle (\Delta \epsilon)^2 \rangle/(2\Delta t) \propto \epsilon^2$; since energy and momentum changes are related by $d\epsilon = v\, dp$, this corresponds to $D_p = D_\epsilon/v^2$, which approaches the GBM scaling $D_p \propto p^2$ (red dashed line) as $v \to \mathbb{C}$ and $\epsilon \to p\,\mathbb{C}$ at fully relativistic momenta. Beyond $p/(m\mathbb{C}) \sim 8$, it transitions to a declining trend approximately following $D_p \propto p^{-1}$.

The regime transition occurs when particle gyroradii exceed the turbulent coherence length. Particles with $r_{\rm c} \gtrsim \Delta_{\rm turb}$ (where $\Delta_{\rm turb}$ is the turbulence coherence length) increasingly sample laminar regions where $\delta B/B_0 \ll 1$. These particles predominantly follow coherent gyromotion with reduced scattering rates, causing the diffusion coefficient to decrease with increasing energy, contrary to the $\propto p^2$ behavior expected for particles confined within the turbulence. The advection coefficient $A_p$ (not shown) similarly increases as $\propto p$ and then saturates at the same transition energy. These trends are consistent with other results reported in this work, including the development of pitch angle anisotropy for high-energy particles (Figure \ref{fig:pitch_dist}) and the inverse scaling of mean energy change $\langle\Delta\epsilon\rangle \propto 1/\epsilon$ (Figure \ref{fig:mean_energy_change}).

This transition reveals that particle acceleration in shear-driven turbulence is intrinsically a multi-scale problem involving the interplay between shear layer width, particle gyroradius, and turbulent coherence length. Although the acceleration mechanism evolves as $r_{\rm c}$ exceeds $\Delta_{\rm turb}$, most particles remain in the regime $r_{\rm c} \lesssim \Delta_{\rm turb}$, confirming that the GBM model and its predicted log-normal distribution accurately describe particles frequently crossing shear layers.

\section{Extrapolation}\label{extrapolation}
\begin{figure}
    \centering
    \includegraphics[width=1.0\linewidth]{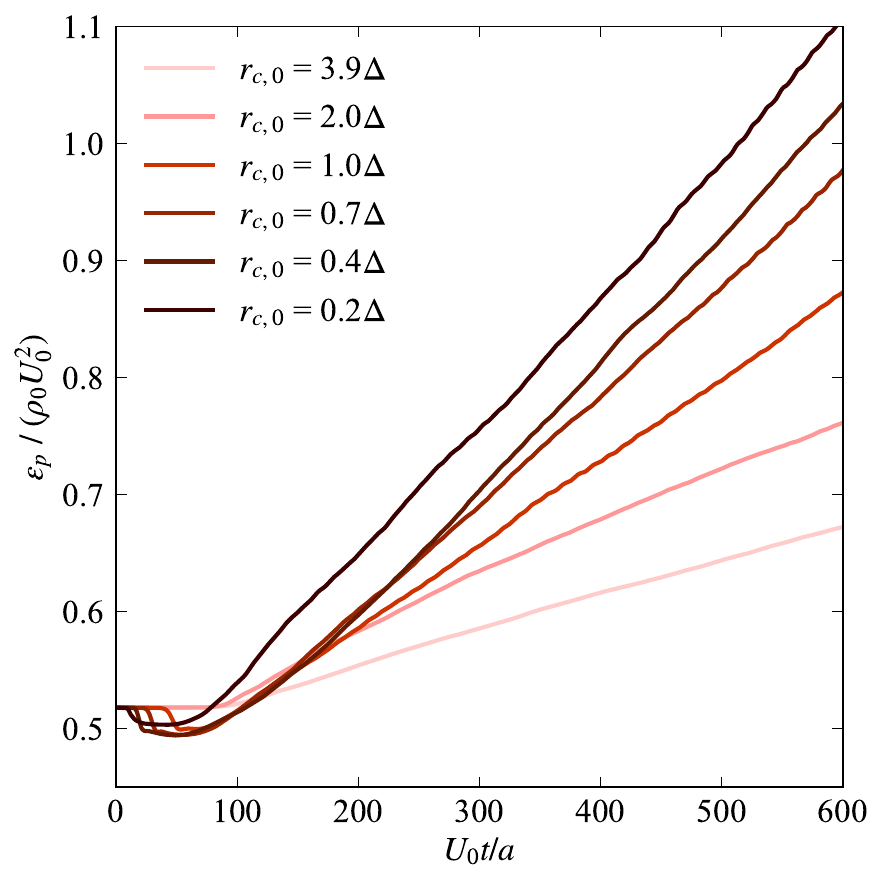}
    \caption{Time evolution of volume-averaged particle energy density $\varepsilon_p$ with varying initial particle gyroradii $r_{\rm c,0}$, controlled by the charge-to-mass ratio $q/m$. All cases exhibit sustained linear energy growth following the turbulence saturation at $t \sim 100\,a/U_0$, verifying that the acceleration mechanism operates across a wide range of gyroradius-to-layer-width ratios.}
    \label{fig:gyroradius}
\end{figure}

Our fiducial simulations initialize particles with gyroradii $r_{\rm c,0}$ approximately equal to the shear layer width, which is defined as $\Delta = 2\,a$. However, this condition is not requisite for the acceleration mechanism reported above. Figure \ref{fig:gyroradius} demonstrates the temporal evolution of particle energy density $\varepsilon_p$ for six different initial gyroradii spanning $0.2\Delta \lesssim r_{\rm c,0} \lesssim 4.0\Delta$. We control the initial gyroradius by varying the charge-to-mass ratio $q/m$ according to $r_{\rm c,0} \propto m /q$, where larger $q/m$ yields smaller $r_{\rm c,0}$.

Two competing effects govern the acceleration efficiency. First, increasing $q/m$ enhances the Lorentz force magnitude $q(\boldsymbol{E} + \boldsymbol{v} \times \boldsymbol{B}/c)$, magnifying particle-field coupling and energy change per scattering. Second, smaller gyroradii reduce the effective scattering cross-section for turbulent interactions, as particles with $r_{\rm c,0} < \Delta$ remain confined to narrow spatial regions and cross shear layers less frequently. Despite these competing effects, Figure \ref{fig:gyroradius} reveals sustained, nearly linear energy growth $\varepsilon_p(t) \propto t$ across all six cases.

This trend across gyroradius space indicates that the acceleration by shear-driven turbulence operates over a wider dynamical range than our fiducial setup explores, a factor of twenty in $r_{c,0}/\Delta$, and is not fine-tuned to $r_{c,0} \sim \Delta$. The runs do not reach the gradual-shear limit $r_c \ll \Delta$, in which the particle mean free path is much smaller than the layer width (Section~\ref{discussion}), so an acceleration history that begins at thermal energies and continues as growing gyroradii exceed the characteristic turbulence scales remains untested here; its bottom-up stage would require the kinetic or hybrid simulations noted in Section~\ref{setup}. Within the tested range, the insensitivity of the mechanism to initial conditions supports its role in the reacceleration of pre-existing energetic particles in shear flows.

\section{Late-Time Turbulence Snapshots}\label{latetime}
\begin{figure}
    \centering
    \includegraphics[width=1.0\linewidth]{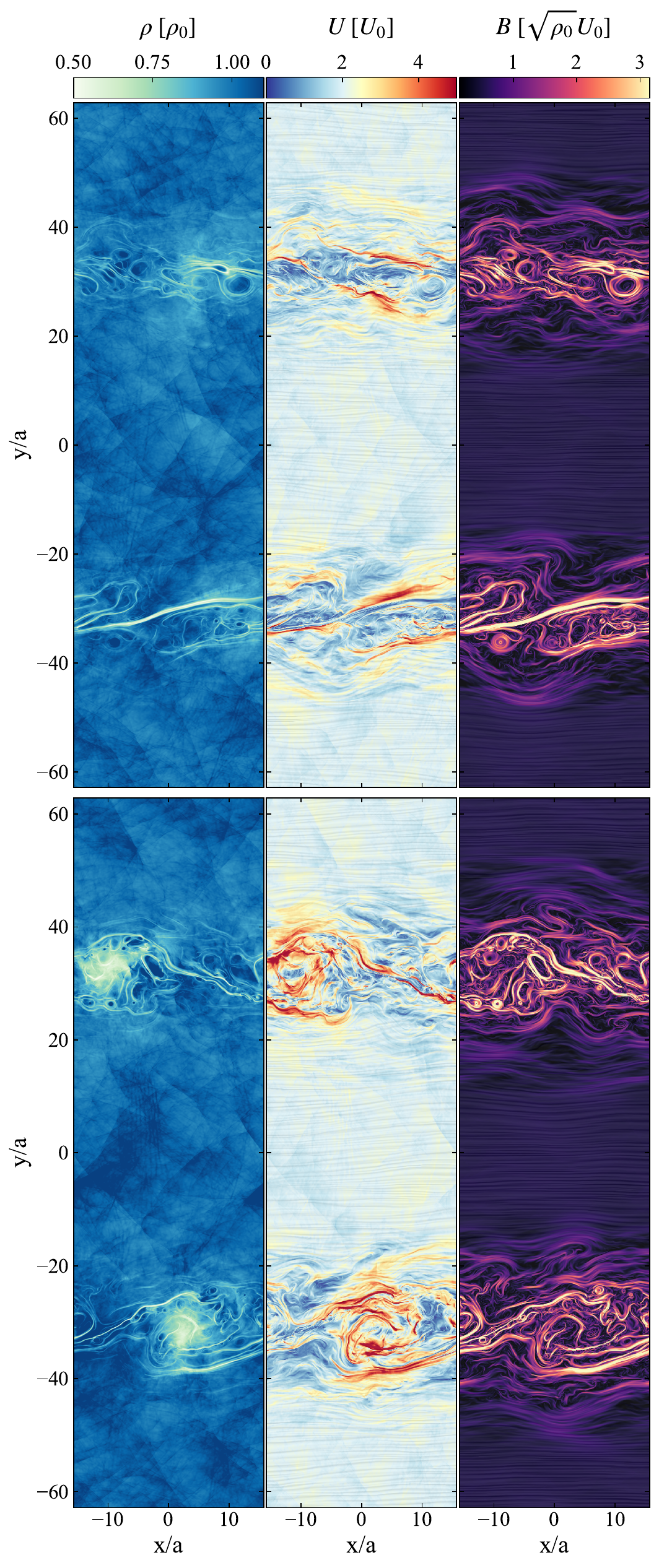}
    \caption{Same as Figure~\ref{fig:KHI}, but at the late times $t = 300$ (top) and $600\,a/U_0$ (bottom), after the turbulence has reached the statistically stationary state. The mixing layers retain similar extent and morphology between the two epochs, differing only in the instantaneous turbulent structures.}
    \label{fig:flow_latetime}
\end{figure}

Figure~\ref{fig:flow_latetime} complements Figure~\ref{fig:KHI} by showing the background fluid density, velocity, and magnetic field at the later times $t = 300$ and $600\,a/U_0$, well after the turbulence reaches the statistically stationary state (Section~\ref{turbulence}). The turbulent mixing layers around the two shear regions retain similar extent and morphology between the two epochs, differing only in the instantaneous realization of the turbulent structures, consistent with the steady mean flow profile (Figure~\ref{fig:shear_profiles}(b)) and the stationary energy balance (Figures~\ref{fig:turbulence} and \ref{fig:steady_state}).

\end{appendix}

\newpage
\bibliography{reference}{}
\bibliographystyle{aasjournalv7}
\end{document}